\documentclass[pdflatex,sn-mathphys]{sn-jnl}% Math and Physical Sciences Reference Style
\RequirePackage{color}
\definecolor{MyDarkGreen}{rgb}{0.02,0.60,0.06}

\usepackage{ulem}
\jyear{2021}%

\theoremstyle{thmstyleone}%
\theoremstyle{thmstyletwo}%

\theoremstyle{thmstylethree}%

\newcommand{\de}[2]{{\frac{\partial #1}{\partial #2}}}

\begin{document}

\title[Warp drive in spherical coordinates with anisotropic matter configurations]{Warp drive solutions in spherical coordinates with anisotropic matter configurations}

%%=============================================================%%
%% Prefix	-> \pfx{Dr}
%% GivenName	-> \fnm{Joergen W.}
%% Particle	-> \spfx{van der} -> surname prefix
%% FamilyName	-> \sur{Ploeg}
%% Suffix	-> \sfx{IV}
%% NatureName	-> \tanm{Poet Laureate} -> Title after name
%% Degrees	-> \dgr{MSc, PhD}
%% \author*[1,2]{\pfx{Dr} \fnm{Joergen W.} \spfx{van der} \sur{Ploeg} \sfx{IV} \tanm{Poet Laureate} 
%%                 \dgr{MSc, PhD}}\email{iauthor@gmail.com}
%%=============================================================%%

\author*[1,2]{\fnm{Gabriel} \sur{Abell\'an}}\email{gabriel.abellan@ciens.ucv.ve}
\email{gabriel@astrumdrive.com}

\author*[1,2,3]{\fnm{Nelson} \sur{Bolivar}}\email{nelsonbolivar@cnea.gob.ar}
\email{nelson@astrumdrive.com}
%\equalcont{These authors contributed equally to this work.}

\author[1]{\fnm{Ivaylo} \sur{Vasilev}}\email{ivaylo@astrumdrive.com}
%\equalcont{These authors contributed equally to this work.}

\affil[1]{\orgdiv{Astrum Drive Technologies}, \orgaddress{\city{Dallas Pkwy Unit 120 B, Frisco, TX.}, \postcode{ 75034},  \country{USA}}}

\affil[2]{\orgdiv{Departamento de F\'isica}, \orgname{Facultad de Ciencias,
Universidad Central de Venezuela}, \orgaddress{\street{Av. Los Ilustres}, \city{Caracas}, \postcode{1041-A},  \country{Venezuela}}}

\affil[3]{\orgdiv{Instituto Balseiro}, \orgname{Centro At{\'{o}}mico Bariloche, Comisi\'on Nacional de Energ\'{\i}a At\'omica (CNEA)--Universidad Nacional de Cuyo (UNCUYO)}, \orgaddress{ \city{Bariloche}, \postcode{8400},  \country{Argentina}}}

%%==================================%%
%% sample for unstructured abstract %%
%%==================================%%

\abstract{
In this work we study the influence of isotropic and anisotropic fluids on the spherically symmetric warp metric. We evaluate the energy conditions and the influence of including a cosmological constant type term. We find that, considering this term, there is a trade--off between the weak and strong energy conditions. The obtained solutions are numerical and we solve the system for both the stationary and the full regime. The influence of imposing the zero expansion condition has been explored. We find a wide diversity of behaviours for the solutions. In general there are regions of spacetime where the energy conditions can be at least partially satisfied. Finally, we calculate the value of the total mass using the density found in the numerical simulations, finding examples where it remains positive during the entire evolution of the system.
}


%%\pacs[JEL Classification]{D8, H51}

%%\pacs[MSC Classification]{35A01, 65L10, 65L12, 65L20, 65L70}

\maketitle

\section{Introduction}\label{intro}
It is known that in general relativity particles can travel globally at superluminal velocities \cite{Morris:1988cz,Visser:1989kh,Krasnikov:1995ad,Everett:1997hb}. Alcubierre explored this idea \cite{Alcubierre:1994tu} and propose a way to drive matter at velocities higher than the speed of light. The mechanism proposed in his work creates a distortion of space-time, called a warp bubble, resulting in a portion of spacetime contracting in front of the bubble and expanding behind of it as the bubble moves through a geodesic. The line element proposed by Alcubierre was,
\begin{equation}
	ds^2 = -dt^2+\left(dx - f(r_s)v_s dt\right)^2 + dy^2+dz^2\;,
\end{equation}
with $r_s = \sqrt{(x-x_s)^2+y^2+z^2}$ and $v_s = \frac{dx_s}{dt}$. This corresponds to an ADM-like decomposition of the line element \cite{Arnowitt:1962hi,Wald1984ucp,Alcub2008itnr}.
Studying this metric, Alcubierre already note that the proposed metric implied the violation of energy conditions, since it seemed that a negative energy density would be necessary to sustain such a bubble.

Since Alcubierre's seminal work, there have been numerous contributions pointing to improve and a better understanding of the physics behind the warp metric he proposed. A key interesting aspect has been to apprehend the properties of spacetimes that would allow superluminal velocities to be achieved \cite{Krasnikov:1995ad,Everett:1997hb}. Among the most relevant contributions are the modifications to the original metric, allowing for a significant decrease in the energy involved \cite{VanDenBroeck:1999sn}. Other explorations implicated the imposition of constraints on the system, either on the spacetime or on the matter fluid, as can be seen in Natario's work \cite{Natario:2001tk} where he proposed a warp drive with zero expansion. Lobo and Visser \cite{Lobo:2004an} also discuss several characteristics of energy-matter of the warp bubble, determining some interesting properties, for instance, it must be massless at the centre. It is worth to remark that in these studies the amount of energy needed is significantly reduced.
One aspect that has received much attention from the very beginning has been the study of energy conditions as a means of validating the physical feasibility of the warp drive \cite{Olum:1998mu,Pfenning:1998ua,Low:1998uy,VanDenBroeck:1999xs,Barcelo:2000zf,Lobo:2002zf,Barcelo:2002bv,Lobo:2004wq,McMonigal:2012ey,Alcubierre:2017kqf,Alcubierre:2017pqm,Lentz:2020euv,Santiago:2021aup,Schuster:2022ati}. The occurrence of horizons and closed time curves as well as the development of instabilities due to the presence of quantum matter has also been studied \cite{Hiscock:1997ya,Clark:1999zn,Gonzalez-Diaz:1999vtr,Eroshenko:2022cyx,Everett:1995nn,Barcelo:2022vru}. This has been the subject of extensive debate in the community and although much progress has been made, there is still no definitive consensus.

In the works mentioned so far, the starting point is the metric and from there the properties of the matter that supports this spacetime are deduced. However, it is possible to proceed in the reverse way, that is to say, by fixing some properties of matter and studying how the corresponding spacetime must be. In a series of papers \cite{Santos-Pereira:2020puq,Santos-Pereira:2021mqp,Santos-Pereira:2021rsr,Santos-Pereira:2021mrr,Santos-Pereira:2021sdm}, Santos et al. propose to study the warp problem from this last point of view. They examined how the elements of the warp metric should be constrained if some kind of matter configuration is imposed, for instance dust or a perfect fluid. In this way they were able to obtain some relations for the deformation function given by the Einstein equations. The present work adheres to this approach. In a previous publication we have proposed an alternative metric to study the issues associated with warp drive \cite{abellan:2023}. This metric exploits the advantages of having spherical symmetry in the description of spacetime.\\ In this work we analyse in depth the consequences on spacetime when considering different types of fluids.%This form serves as an alternative to the way the problem has been treated in early works.

In section \ref{sec2} we discuss the most important results related to the warp metric in spherical coordinates. We also present the energy-momentum tensor that we will use to describe the type of fluids we will consider. On the other hand we will give a description of the system of equations obtained in terms of traveling patterns. This allows us to study the stationary behavior of the system under study. Section \ref{sec3} is devoted to evaluate the energy conditions associated with the problem in question. This will be done from both the fluid and metric points of view. We will then look at the consequences to the system after imposing the zero expansion condition. Subsequently, in sections \ref{sec4} and \ref{sec5} we will go on to solve numerically the system of equations obtained. Basically we will study solutions for isotropic fluids and then for anisotropic fluids. The solutions studied will be both in the steady state and in the full regime. We also include a discussion of the effect of adding a cosmological constant type term to the system. In section \ref{sec6} we will use the results of the simulations to determine the behavior of the mass as the system evolves. Finally we make some concluding remarks and considerations for future work.

\section{Warp drive in spherical coordinates}\label{sec2}
In this section we present the ingredients of the model we intend to build. We want to write Einstein's equations for the metric proposed. That is
\begin{eqnarray}
    G_{\mu\nu} \equiv R_{\mu\nu} - \frac{1}{2}Rg_{\mu\nu} = 8\pi T_{\mu\nu}\;.
\end{eqnarray}
Here $R_{\mu\nu}$ is the Ricci tensor, $R$ the Ricci scalar, $g_{\mu\nu}$ the metric and $T_{\mu\nu}$ the energy--momentum tensor. Also we are working in geometric units where $G=c=1$.

\subsection{Metric and Einstein Tensor}
In a recent work \cite{abellan:2023}, motivated by Alcubierre's work we propose a warp line element using spherical coordinates
\begin{eqnarray}\label{metric}
    ds^2 = -dt^2 + (dr - \beta dt)^2 + r^2d\Omega^2\;,
%    -(1 - \beta^2)dt^2 - 2\beta dt dr + dr^2 + r^2d\Omega^2\;,
\end{eqnarray}
with $d\Omega^2 = d\theta^2 + \sin^2\theta d\phi^2$ and $\beta$ the form function. Note the non--diagonal element and the fact that $\beta=0$ returns a Minkowski flat spacetime.

Using this metric we calculate the elements of the Einstein tensor, which are
\begin{eqnarray}
	G_{00} &=& \frac{\beta}{r^2}\Bigg[\left(1-\beta^2\right)\left(\beta+2r \frac{\partial \beta}{\partial r}\right)
	- 2r \beta \frac{\partial \beta}{\partial t}\Bigg], \label{t-ei-1} \\
	G_{01} &=& \frac{\beta}{r^2}\left(\beta^2 + 2r \beta \frac{\partial \beta}{\partial r}+2 r \frac{\partial \beta}{\partial t}\right) , \label{t-ei-2} \\
	G_{11} &=& -\frac{1}{r^2}\left(\beta^2 + 2r \beta \frac{\partial \beta }{\partial r}+ 2 r \frac{\partial \beta}{\partial t}\right) , \label{t-ei-3}\\
	G_{22} &=& - r\Bigg\{ \beta \left(2 \frac{\partial \beta}{\partial r}+r \frac{\partial^2 \beta}{\partial r^2}\right)+\frac{\partial \beta}{\partial t} % \nonumber\\
	+ \;r \left[\left(\frac{\partial \beta}{\partial r}\right)^2+ \frac{\partial^2 \beta}{\partial t \partial r}\right] \Bigg\} , \label{t-ei-4} \\
	G_{33}&=& - r\sin^2{\theta}\Bigg\{ \beta \left(2 \frac{\partial \beta}{\partial r}+r \frac{\partial^2 \beta}{\partial r^2}\right) + \frac{\partial \beta}{\partial t}
	 + r \left[\left(\frac{\partial \beta}{\partial r}\right)^2+ \frac{\partial^2 \beta}{\partial t \partial r}\right] \Bigg\} \label{t-ei-5} .
\end{eqnarray}
In general, we consider the form function as a quantity dependent of both time and radial coordinates. Note also that we only have five non--zero components. This represents a remarkable simplification with respect to the Alcubierre's metric written in Cartesian coordinates.

\subsection{Energy--Momentum tensor}
The next step in writing the Einstein equations consists in fixing the material content of the system under study. For this we consider an anisotropic fluid given by
\begin{equation}\label{temunuani}
	T_{\mu \nu}^{(a)}=(\rho + p_\perp)u_\mu u_\nu + p_\perp g_{\mu\nu} + (p_r - p_\perp)s_\mu s_\nu \;,
\end{equation}
where we have used the following four--vectors
\begin{equation}\label{4vectors}
    u_\mu = (-1,0,0,0)\;, \hspace{.7cm}
    s_\mu = (-\beta,1,0,0)\;,
\end{equation}
these vectors are timelike and spacelike respectively and satisfy the relations $u^\mu u_\mu = -1$, $s^\mu s_\mu = 1$ and $u^\mu s_\mu = 0$. In (\ref{temunuani}) $\rho$ is the relativistic energy density, $p_r$ the radial pressure and $p_\perp$ the tangential pressure.

We are interested in considering the effects of including a cosmological constant type term. To facilitate the subsequent numerical solution of the system of equations, we will include this term in the energy--momentum tensor, namely
\begin{equation}\label{temunucosmo}
    T_{\mu\nu}^{(\Lambda)}= -\frac{\Lambda}{8\pi}g_{\mu\nu}
    = -\rho_{\Lambda} g_{\mu\nu}\;,
\end{equation}
with $\Lambda$ (and $\rho_\Lambda$) a constant.

So, the total energy--momentum tensor is $T_{\mu\nu} \!=\! T_{\mu\nu}^{(a)} + T_{\mu\nu}^{(\Lambda)}$. Using equations (\ref{temunuani}) and (\ref{temunucosmo}) we can write in matrix form
%\begin{widetext}
\begin{equation}\label{temunumatrix}
	T_{\mu\nu} = \left[ \begin{array}{cccc}
		\rho + \beta^2 p_r + (1-\beta)\rho_{\Lambda} & -\beta (p_r - \rho_{\Lambda}) & 0 & 0\\
		-\beta (p_r - \rho_{\Lambda}) & p_r - \rho_{\Lambda} & 0 & 0\\
		0 & 0 & r^2 (p_\perp - \rho_{\Lambda}) & 0 \\
		0 & 0 & 0 & r^2 \sin^2\!\theta\, (p_\perp - \rho_{\Lambda})
	\end{array} \right] .
\end{equation}
%\end{widetext}
It is interesting to compare with the tensor proposed in \cite{Santos-Pereira:2021sdm}. We can easily check that the system admits anisotropic configurations. 

\subsection{Einstein's equations}
Using the Einstein tensor components (\ref{t-ei-1})--(\ref{t-ei-5}) and the energy--momentum tensor (\ref{temunumatrix}) we find the Einstein's equations for this system
\begin{eqnarray} 
    \beta \left(\beta+2r \de{\beta}{r}\right) &=& 8\pi r^2 \rho^{(e)}\,, \label{eineq1} \\
    \beta\left(\beta + 2r\de{\beta}{r}\right) + 2r \frac{\partial\beta}{\partial t}  &=&  -8\pi r^2 p_r^{(e)} , \;\;\;\;\; \label{eineq2} \\
	\beta^2 + r \de{\beta}{t} - r^2 \left[\de{}{r}\left(\beta \de{\beta}{r}\right) + \frac{\partial^2\beta}{\partial t \partial r}\right] &=& 8\pi r^2 \Delta \,, \label{eineq3}
\end{eqnarray}
with the effective quantities $\rho^{(e)}=\rho + \rho_\Lambda$, $p_r^{(e)} = p_r - \rho_\Lambda$ and $\Delta = p_\perp - p_r$ the anisotropy factor. From Einstein tensor (\ref{t-ei-2}) and (\ref{t-ei-3}) the equation (\ref{eineq2}) arise. The same is true for the components (\ref{t-ei-4}) and (\ref{t-ei-5}) which produce the same equation for $p_\perp^{(e)}=p_\perp - \rho_\Lambda$. That is, we have just three independent equations to solve. The system (\ref{eineq1})--(\ref{eineq3}) has four degrees of freedom: $\beta$, $\rho$, $p_r$ and $\Delta$.

\subsection{Stationary, traveling--type solutions}
We are interested in the possibility that the system of equations produces a traveling wave type solution. For this we use the parametrization
\begin{equation}
    \beta(t,r) = f(r_s) v_s(t)\;.
\end{equation}
Here $r_s = \|r - v_s(t) t\|$, with $v_s(t)$ a time--dependent spatial velocity function. In the following we consider this velocity a constant, so the displacement is uniform in radial direction.

Using this parametrization we find that the equations to be solved for $r - v_s t \geq 0$ are
%\begin{widetext}
\begin{eqnarray}
     f^2 + 2(r_s+v_s t) f f' = \frac{8\pi}{v_s^2}(r_s+v_s t)^2 \rho^{(e)} \;, \label{stat1} \\
f^2 + 2(r_s+v_s t)(f-1)f' =  -\frac{8\pi}{v_s^2}(r_s+v_s t)^2 p_r^{(e)}\;, \label{stat2} \\
 f^2 - (r_s+v_s t)f' - 
 (r_s+v_s t)^2[(f')^2+(f-1)f'']  = \frac{8\pi}{v_s^2}(r_s+v_s t)^2 \Delta\,, \label{stat3}
\end{eqnarray}
%\end{widetext}
where primes denote derivatives with respect to $r_s$. Moreover, for $r - v_s t < 0$ the system vanish and we obtain the traveling wave behavior.

The systems of equations (\ref{eineq1})-(\ref{eineq3}) and (\ref{stat1})-(\ref{stat3}) describe the complete and stationary system respectively. To solve it is necessary to provide boundary conditions. In both cases we will use
\begin{equation}
    \lim_{r\to 0} \beta = \lim_{r\to \infty} \beta = 0\;.
\end{equation}
An important point to note is that only the complete system requires initial conditions. This is one of the advantages of working with the equations that explore travelling patterns.

\section{Energy conditions and expansion}\label{sec3}
In this section we study energy conditions, which are constraints imposed on the energy-momentum tensor so that we can evaluate the viability of the model and control non-physical aspects of the system \cite{Wald1984ucp,Visser1995lwet,Poisson2004rtmb,Carroll2004sgig,Curiel:2014zba,Kontou:2020bta}.

\subsection{Weak energy condition}
Weak energy condition (WEC) requires that $u^\mu$, $T_{\mu \nu} u^\mu u^\nu \geq 0$. That is
\begin{equation}\label{wec}
	 T_{\mu \nu} u^\mu u^\nu = \rho + \rho_\Lambda \geq 0 \;.
\end{equation}
Here $u_\mu$ is given by (\ref{4vectors}) and is a timelike vector. In fact, this vector defines an Eulerian observer for the system. So, we find that weak energy condition is satisfied if $\rho + \rho_\Lambda \geq 0$.
Using Einstein's equations, the equivalent condition on the $G_{\mu\nu}$ requires that
\begin{equation}
     \beta \left(\beta + 2r \de{\beta}{r}\right)  \geq 0\;.
\end{equation}
For this result we have used $r^2>0$, which simplifies the final expression.

\subsection{Dominant energy condition}
The dominant energy condition (DEC) is equivalent to the WEC, with the additional requirement that $T^{\mu}_{\nu} u^\nu$ is a future-pointing causal vector. Thus, the weak energy condition and $F^\mu F_\mu \leq 0$, with $F^\mu = T^{\mu \nu}u_\nu$, must be satisfied.

After some straightforward calculations, we find that
\begin{equation}\label{dec}
    F^\mu F_\mu = -(\rho+ \rho_\Lambda)^2 \leq 0 \;,
\end{equation}

Note that, if the conservation of the $T_{\mu \nu}$ is also considered, both conditions (\ref{wec}) and (\ref{dec}) guarantee the causal structure in local matter configurations. In terms of the metric components,
the dominant condition is equivalent to

\begin{equation}
    - \left( \beta + 2 r\frac{\partial\beta}{\partial r} \right)^2 \leq 0 \;.
\end{equation}
It is clear that this condition is always satisfied.

\subsection{Strong energy condition}
The strong energy condition (SEC) imposes a bound on a more complicated expression in $4$ dimensions
\begin{equation}
	\left(T_{\mu \nu}-\frac{1}{2}Tg_{\mu\nu}\right)u^\mu u^\nu \geq 0 \,.
\end{equation}
Using the energy--momentum tensor (\ref{temunumatrix}) and the definition for the anisotropy factor $\Delta$ we obtain
\begin{equation}\label{sec}
%	\rho - 2\rho_\Lambda + p_r + 2p_\perp =
    \rho - 2\rho_\Lambda + 3p_r + 2\Delta
    \geq 0 \,.
\end{equation}
This is a proper generalization of results in \cite{Santos-Pereira:2021sdm}.
The equivalent form written in terms of $\beta$ is

\begin{equation}
   -\beta \left(2 \frac{\partial \beta}{\partial r}+r \frac{\partial^2 \beta}{\partial r^2}\right)+2 \frac{\partial \beta}{\partial t}+r \left[\left(\frac{\partial \beta}{\partial r}\right)^2+\frac{\partial^2 \beta}{\partial t \partial r}\right] \geq 0\;.
\end{equation}
This expression is actually rather complex in relation to the others already discussed.

\subsection{Null energy condition}
Finally, the null energy condition (NEC) is analogous to WEC, with the timelike vector $u_\mu$ replaced by a null vector $k_\mu$. That is
\begin{equation}\label{NEC}
    	T_{\mu \nu }k^\mu k^\nu \geq 0\,.
\end{equation}
Assuming a vector $k^\mu = \{a_\pm,b,0,0\}$, with 
\begin{equation}
    a_\pm = \frac{b}{\beta \pm 1}\;,
\end{equation}
which is a lightlike vector $k^\mu k_\mu=0$. We find that NEC is
\begin{eqnarray}
	\rho + p_r \geq 0 \;.
\end{eqnarray}
Note that the contributions from $\rho_\Lambda$ cancel each other. Writing this condition in terms of $\beta$ we find
\begin{equation}
    \frac{\partial \beta}{\partial t} \leq 0\;. \label{null_condition_beta}
\end{equation}
That is, as time evolves, $\beta$ decreases.

Finally a comment about the cosmological constant type term. It is important to note that the addition of $\rho_\Lambda$ in the model gives additional freedom, allowing a shift to be made in case of finding conditions that are not satisfied by themselves. We consider this to be important because it can help to establish more general procedures to deal with the problems that warp drive has traditionally suffered with respect to energy conditions. However, we can see that there is an interplay between the weak and strong conditions. That is, we can manipulate the term and enforce one of them but necessarily make the other one worse. We will see examples of this when studying the solutions.

\subsection{Expansion}
\begin{center}
    \begin{figure}
    \centering
    \includegraphics[scale=0.3]{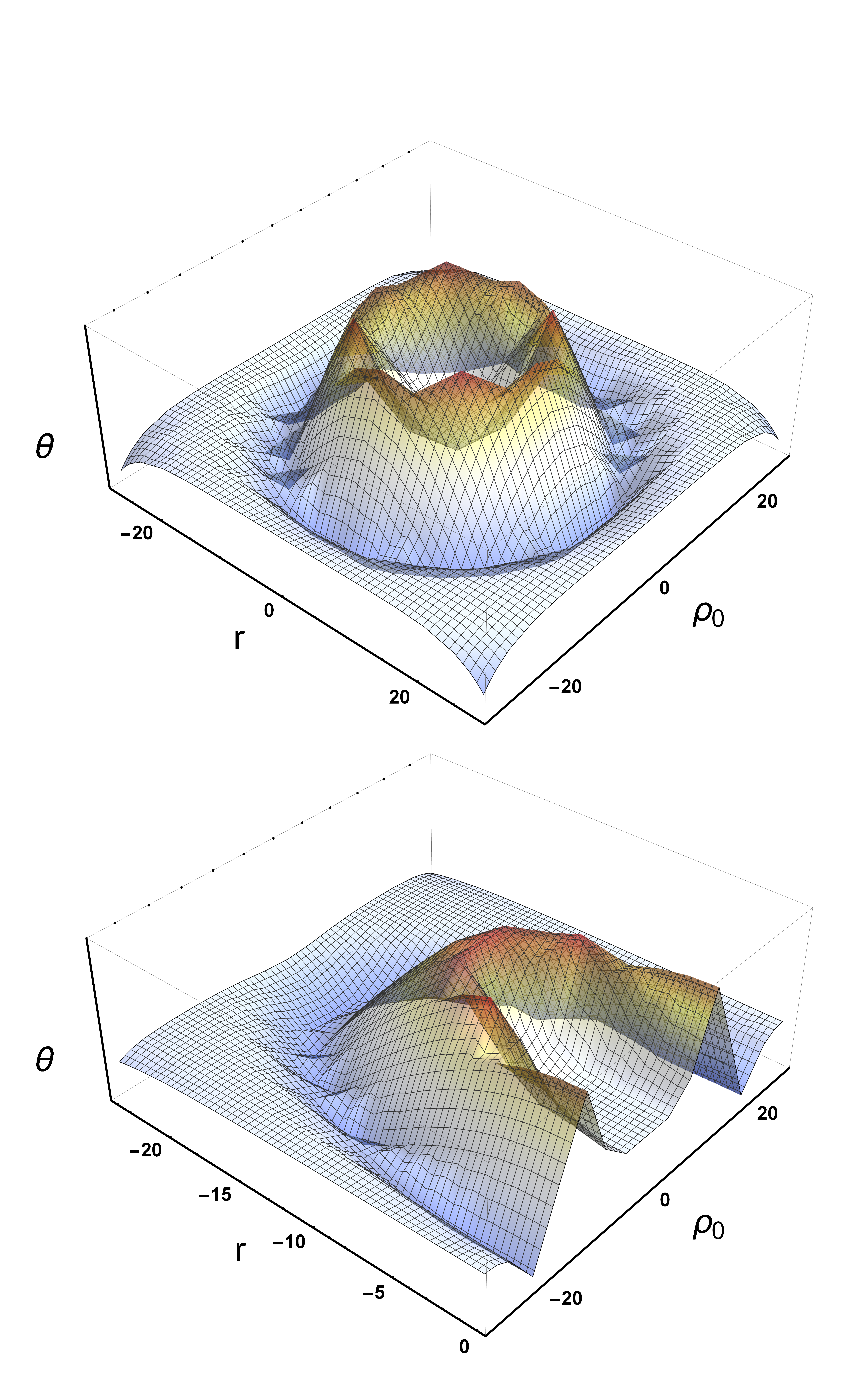}
    \caption{The plot depicts the expansion of the volume elements $\Theta$ of the spherical warp bubble in the non-stationary case with $\rho_0$ being a form of parametrization to account for the coordinates ``perpendicular to $r$''. The full expansion is shown in the upper plot. A transverse cut is shown in the lower plot where it can be noted the flat sitting region and the expansion and contraction.}
    \label{fig:expansion}
\end{figure}
\end{center}

As can be seen in (\ref{metric}), the geometry corresponding to 3-space is flat and therefore the extrinsic curvature tensor $K_{ij}$ allows encoding information about the curvature. The extrinsic curvature tensor serves to describe how the 3-dimensional hypersurfaces are embedded in the 4-dimensional space. It is given by the expression
\begin{equation}
    K_{ij} = \frac{1}{2\alpha} \Big(
    D_{i}\beta_j + D_{j}\beta_i - \frac{\partial g_{ij}}{\partial t}\Big)\;,
\end{equation}
where $D_i$ is the covariant derivative relative to the 3-space metric, and $\alpha$ correspond to lapse function which we assume $\alpha = 1$ as usual. This choice means that timelike curves normal to the hypersurfaces in 3-space are geodesics, namely Eulerian observers are free falling. According to these observers, the expansion $\Theta$ of the volume elements is written in terms of $K_{ij}$ and is given by
\begin{eqnarray}\label{expansion}
    \Theta &=& - \mathrm{Tr}[K_{ij}] \nonumber\\
   &=& 2\frac{\beta}{r} + \frac{\partial\beta}{\partial r} \;.
\end{eqnarray}
In fig. \ref{fig:expansion} is possible to observe the behaviour of expansion for a typical $\beta$ solution (see next sections). Analogously to Alcubierre's findings, there is a deformation that lives in a compact domain. The volume elements are expanding behind and contracting in a sort of wavefront. We also note the existence of a flat region inside the bubble. All this will have effects on the dynamical equations and energy conditions as we will see next.

It is possible to use equation (\ref{expansion}) and impose a zero expansion condition on the Einstein equations. This restricts the possibilities of fluid evolution and hence the shape of spacetime. With $\Theta = 0$, we find that
\begin{eqnarray}
    \beta^2 &=& -\frac{8\pi}{3}r^2 \rho\;, \nonumber \\
    \frac{2}{r} \frac{\partial\beta}{\partial t} -
    \frac{3}{r^2}\beta^2 &=& -8\pi p_r\;, \\
    \frac{3}{r} \frac{\partial\beta}{\partial t}
    - \frac{9}{r^2}\beta^2 
    - \frac{\beta}{r^2}\frac{dr}{dt} &=& 8\pi\Delta\;.
\end{eqnarray}
These equations are interesting because it is evident that in order to have a $\beta$ real, $\rho$ must be negative. So we have a close relation between the expansion factor and the energy density as was already pointed out in \cite{Natario:2001tk}.

To conclude this section we find it is worthwhile to evaluate the energy conditions using the null expansion condition. Using the constraint $\Theta = 0$ over expression (\ref{expansion}) we find in terms of $\beta$ that
\begin{eqnarray}
    \mbox{WEC:} & \hspace{.6cm} \beta^2 \leq 0\;, \\
    \mbox{DEC:} & \hspace{.6cm} \beta^2 \geq 0\;, \\
    \mbox{SEC:} & \hspace{.6cm} \beta^2 \geq 0\;, \\
    \mbox{NEC:} & \hspace{.6cm} \displaystyle \frac{\partial\beta}{\partial t} \leq 0\;.
\end{eqnarray}
It is clear that the weak condition cannot be satisfied under the zero expansion regime. However, the dominant and strong conditions are always satisfied. The null condition does not undergo any modification using the constraint.

\section{Case 1: isotropic solutions}\label{sec4}
Since the system of equations is highly non-linear, we will solve it numerically. In order to do so, we consider several cases separately.

\subsection{Non--stationary solutions ($\beta(t,r)$)}
First we solve for system (\ref{eineq1})--(\ref{eineq3}) without imposing any constraint. In this case we set $\Delta=0$ so we reduce de degrees of freedom to three and the system closes. Also we set $\rho_\Lambda = 0$ and later we consider the effects when this is included.
In order to solve it is necessary to give an initial condition for $\beta$. We will use for convenience a Gaussian given by
\begin{equation}\label{inicond}
    \beta(0,r) = \exp{\Big[-\frac{(r-0.5r_e)^2}{0.25r_e}\Big]}\;.
\end{equation}
Here $r_e$ represents the spatial maximum value for the numerical integration domain.

\begin{center}
    \begin{figure*}
    \centering
    \includegraphics[scale=.8]{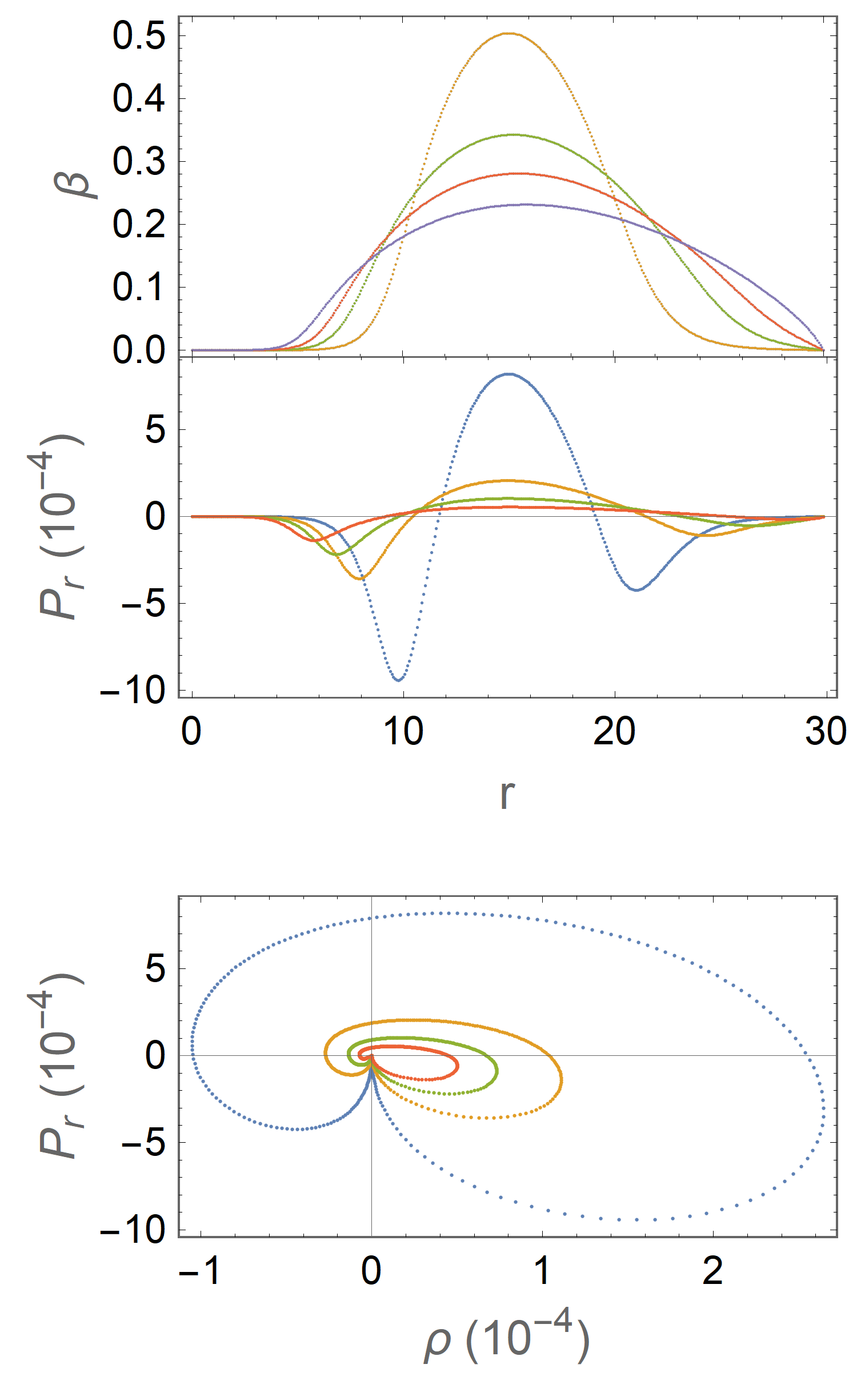}
    \caption{Non--stationary Case 1. Form function $\beta$, radial pressure $p_r$ and empirical equation of state for $t=1$ (blue), $3$ (orange), $5$ (green) and $8$ (red).}
    \label{fig:non-sta-case1-1}
\end{figure*}
\end{center}

In figure \ref{fig:non-sta-case1-1} we have the form function $\beta$, radial pressure $p_r$ and the empirical equation of state $p_r(\rho)$ for several times. We can notice that $\beta$ is decreasing in amplitude as it becomes wider. We also see that the radial pressure supporting this warp bubble has regions where it is positive and others where it is negative. This is interesting because it tells us that this distribution of matter generates non-zero pressure gradients that could affect the motion of the system. On the other hand, empirical equation of state shows a completely non-linear behaviour and is multivalued.

\begin{center}
    \begin{figure*}
    \centering
    \includegraphics[scale=0.8]{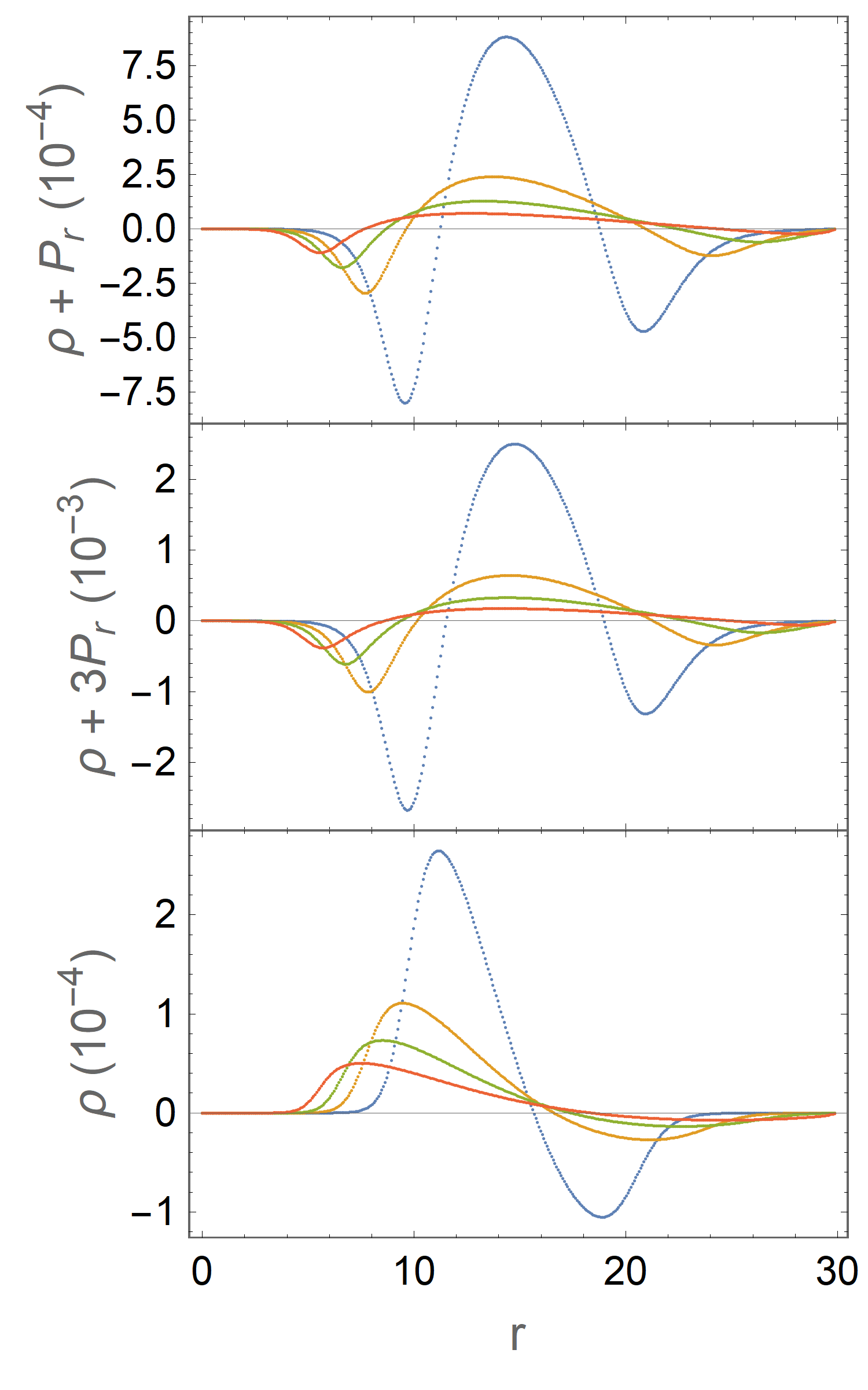}
    \caption{Non--stationary case 1. From the top we show NEC, SEC and WEC for $t=1$ (blue), $t=3$ (orange), $t=5$ (green) and $t=8$ (red).}
    \label{fig:non-sta-case1-2}
\end{figure*}
\end{center}

Below in the graph \ref{fig:non-sta-case1-2} we have the energy conditions. In general we can observe that for all of them there are regions where they are fulfilled and others where they are not. However, we notice that for all the cases they are bounded and therefore they do not have singular behaviour. This is relevant since it allows us to use the cosmological constant term to enforce some of them. However, we see the interplay between the strong and weak conditions as we have already mentioned. We also find that the null condition is violated in some regions as has been suggested in \cite{Santiago:2021aup}.

\subsection{Stationary solutions ($v_s f(r_s)$)}
\begin{center}
    \begin{figure*}
    \centering
    \includegraphics[scale=.65]{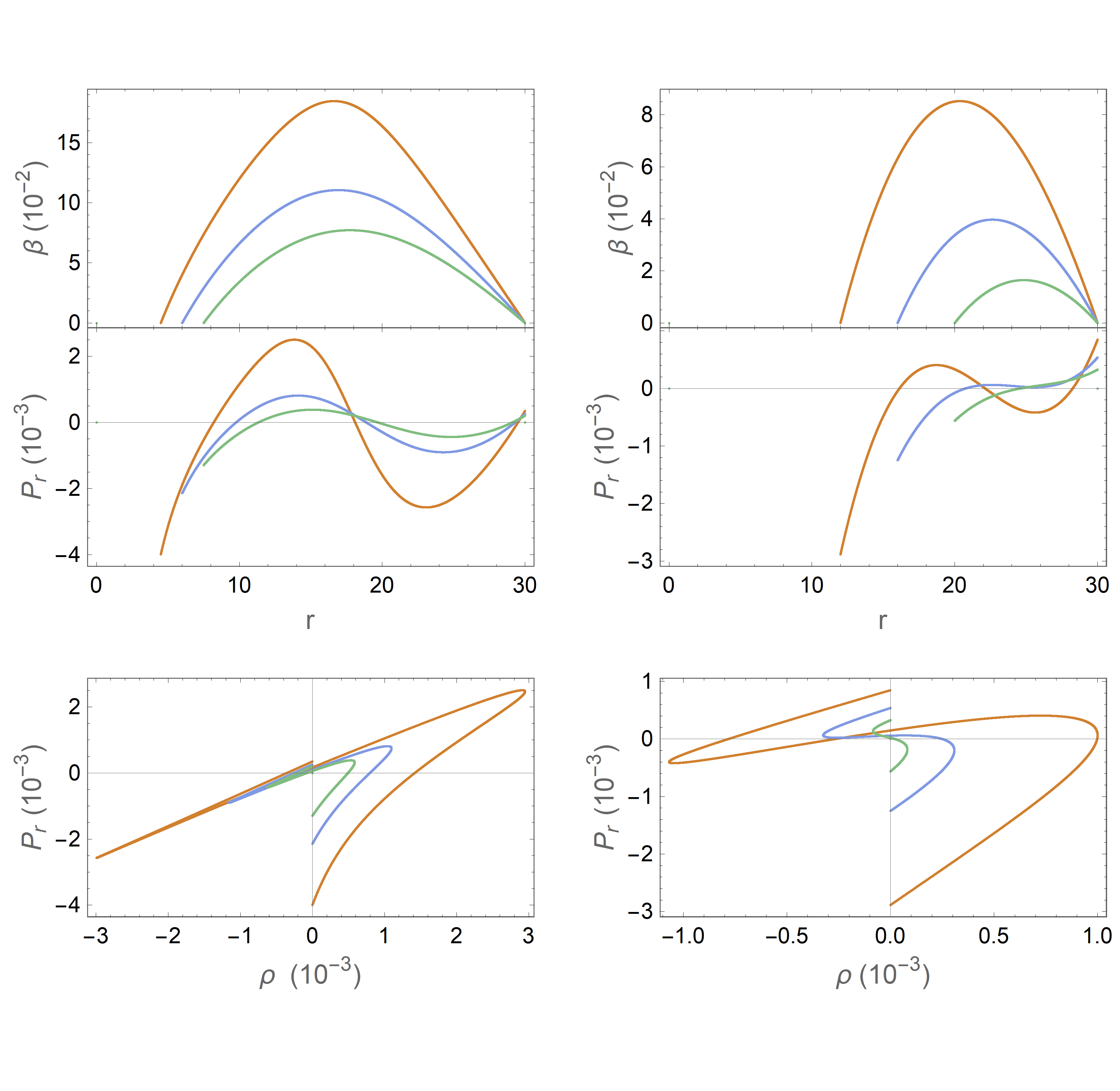}
    \caption{Stationary case 1 for $v_s=0.3$ (left column) and $v_s=0.8$ (right column). Each curve corresponds to times $t=15$ (orange), $t=20$ (blue) and $t=25$ (green)}
    \label{fig:sta-case1-1}
\end{figure*}
\end{center}

\begin{center}
    \begin{figure*}
    \centering
    \includegraphics[scale=0.65]{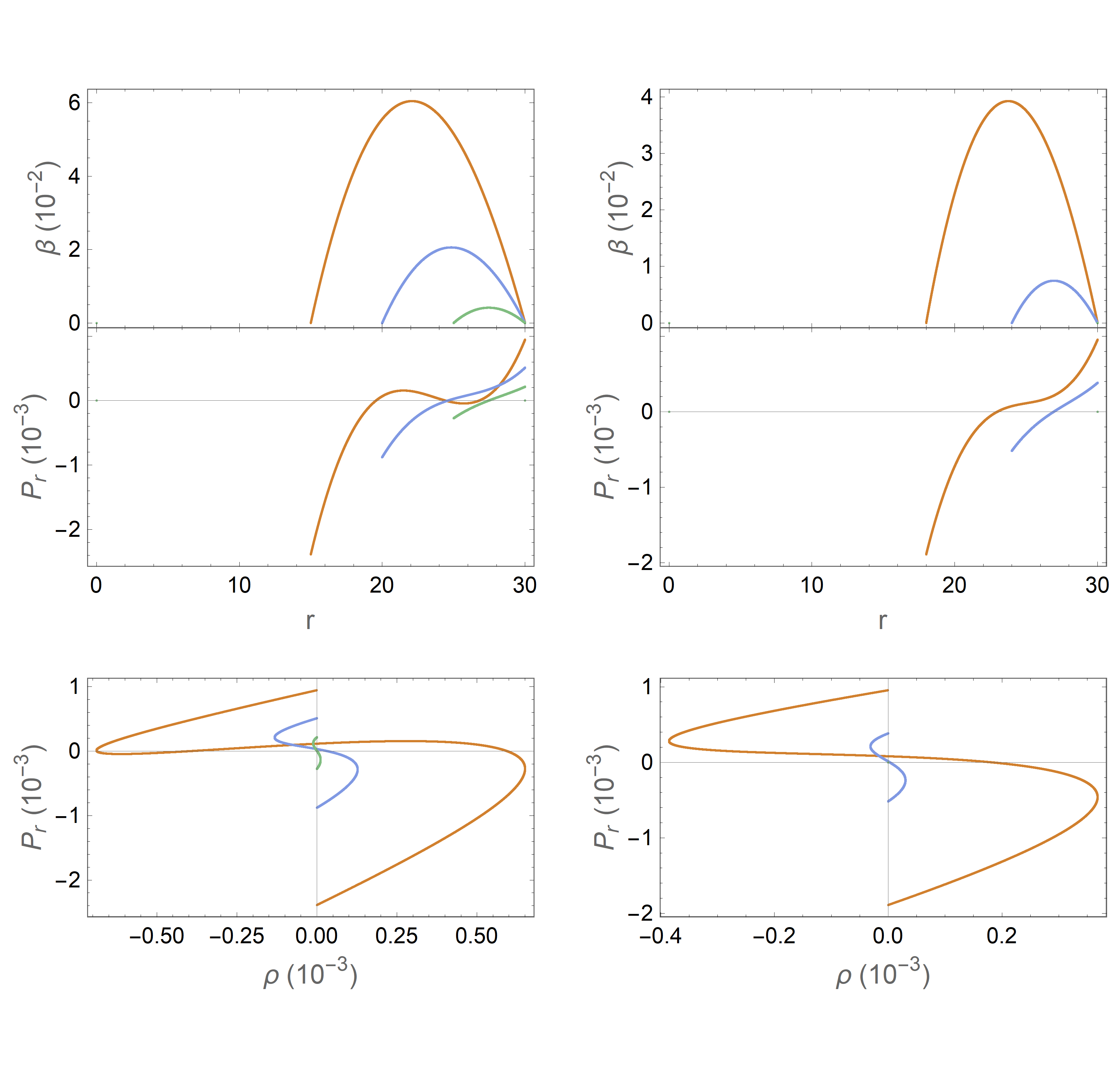}
    \caption{Stationary case 1 for $v_s=1.0$ (left column) and $v_s=1,2$ (right column). Each curve corresponds to times $t=15$ (orange), $t=20$ (blue) and $t=25$ (green)}
    \label{fig:sta-case1-2}
\end{figure*}
\end{center}

The traveling wave type solutions are interesting because cover the purpose of what was originally proposed by Alcubierre. They also represent a sort of stationary behavior for the system under study.
In fig. \ref{fig:sta-case1-1} and fig. \ref{fig:sta-case1-2} we show the stationary solutions of $\beta$, pressure $p_r$ and the empirical equation of state $p_r(\rho)$. These are shown for different velocities $v_s$, ranging from $v_s=0.3$ (subluminal) to $v_s=1.2$ (superluminal).
In fig. \ref{fig:sta-case1-1} velocities $v_s=0.3$ and $v_s=0.8$ are depicted. As expected the amplitude of $\beta$ decreases with time, although the lower velocity shows the largest initial amplitudes. This behavior is also seen in the superluminal case, fig. \ref{fig:sta-case1-2}. 
The form function $\beta$ displaces its origin with the velocity, the bubble shape shrinks with larger $v_s$. 

The radial pressure as in the non--stationary case has positive and negative regions constrained to the size of $\beta$ for each time and velocity. Nevertheless, the distribution of the pressures differs from the non--stationary case in what we think is an interesting feature. For instance, the boundary values of the pressure that sustains the bubble are finite, negative at lower $r$ (the origin point of the $\beta$ form), and positive by the external boundary. This is in contrast with the zero boundary in non-stationary scenarios. It is worth noting then that the pressure distribution suggests, for larger subluminal velocities a pressure gradient along the bubble. 

Looking at the empirical equations of state, one can notice the appearance of regions where $P_r$ can be linearly approximated. Nevertheless, we note, as with the non-stationary solutions, the fact that the relationship between $P_r$ and $\rho$ is multivalued.\\
We can divide the empirical plots into quadrants, and then observe that the system possesses regions with different combinations for signs. We want to remark the upper right quadrant which shows positive signs for $\rho$ and $P_r$.
These positive regions has larger contributions for times chosen when $v_s=0.3$, but as the velocity increases the contributions for larger times are smaller, as can be seen in fig. \ref{fig:sta-case1-1} and even more pronounced in fig. \ref{fig:sta-case1-2} where only shorter times has both positive contributions.

\begin{center}
    \begin{figure*}
    \centering
    \includegraphics[scale=.65]{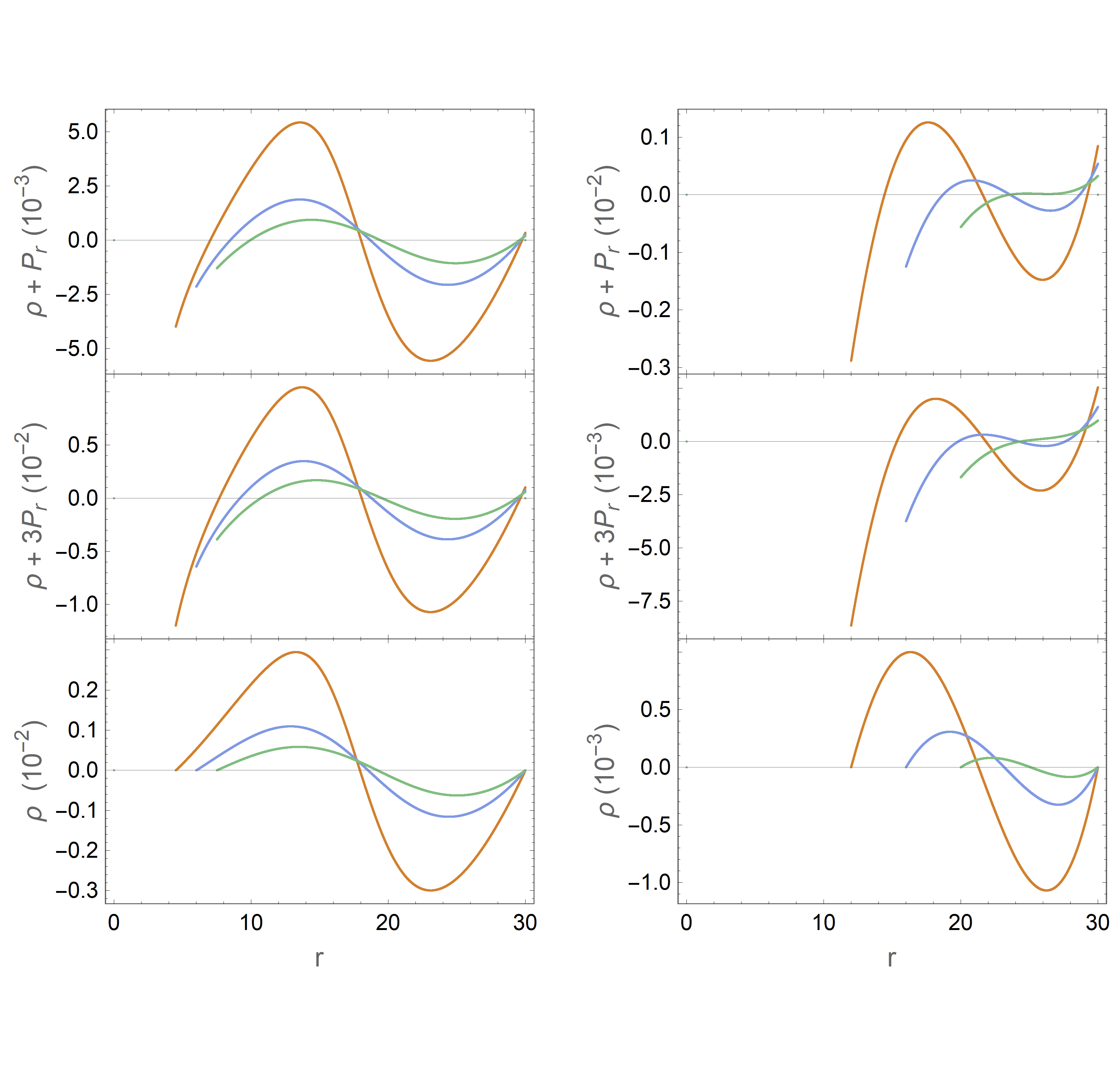}
    \caption{Stationary case 1. From the top down we show the NEC, SEC and WEC for $t=15$ (orange), $t=20$ (blue) and $t=25$ (green). Left column $v_s=0.3$ and right column $v_s=0.8$.}
    \label{fig:sta-case1-ec1}
\end{figure*}
\end{center}

\begin{center}
    \begin{figure*}
    \centering
    \includegraphics[scale=.65]{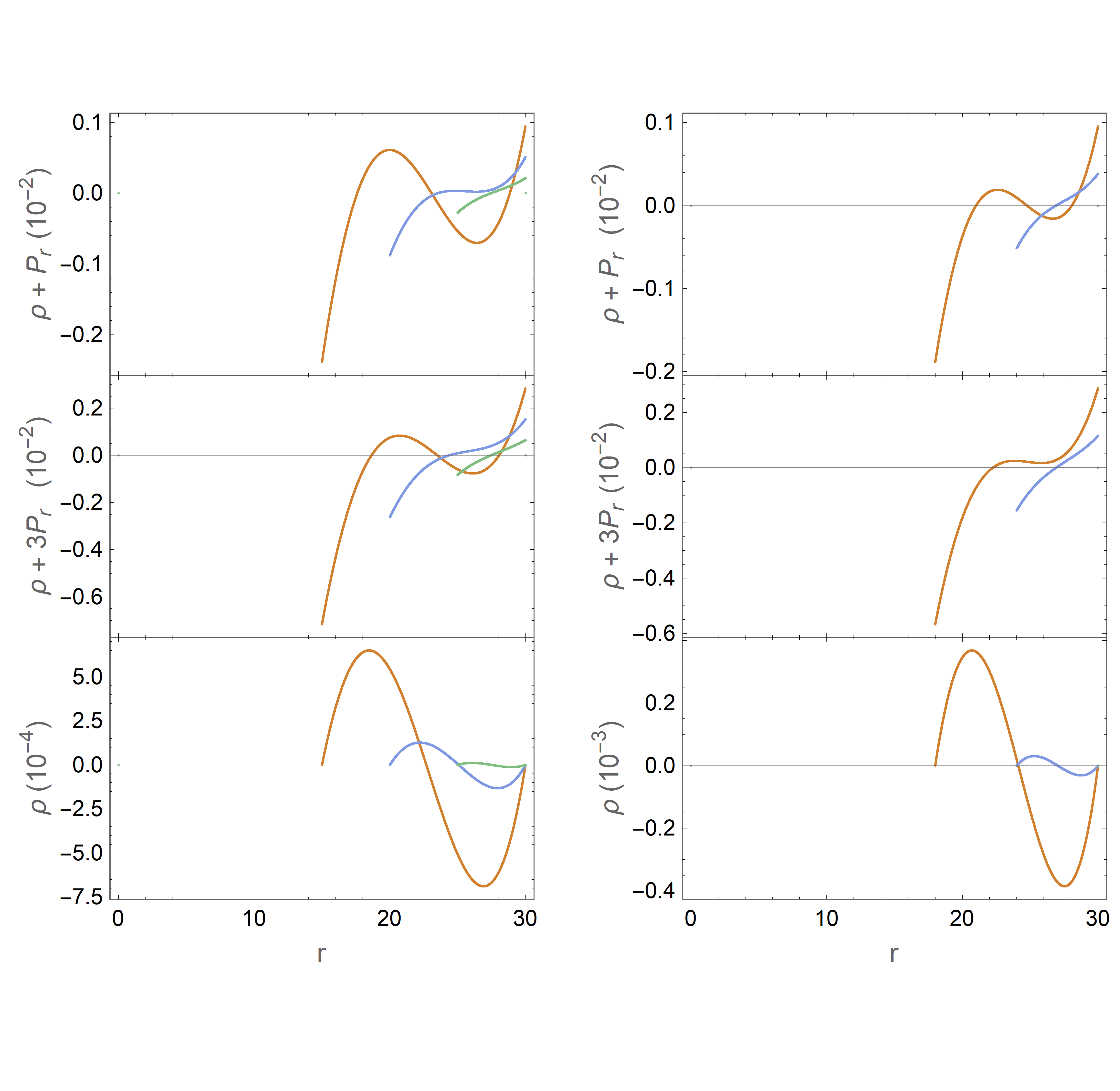}
    \caption{Stationary case 1. From the top down we show the NEC, SEC and WEC for $t=15$ (orange), $t=20$ (blue) and $t=25$ (green). Left column $v_s=1.0$ and right column $v_s=1.2$.}
    \label{fig:sta-case1-ec2}
\end{figure*}
\end{center}

\section{Case 2: anisotropic solutions}\label{sec5}
In this case we consider an arbitrary anisotropy factor $\Delta$, so it is necessary to give an additional condition to close the system. 

First we will give an equation of state in the extended polytrope form
\begin{equation}\label{eos}
    p_r = K\rho^\gamma + \epsilon \rho\;,
\end{equation}
with $K$, $\gamma$ and $\epsilon$ parameters that need to be provided.

\subsection{Non--stationary solutions ($\beta(t,r)$)}
\begin{center}
    \begin{figure*}
    \centering
    \includegraphics[scale=0.65]{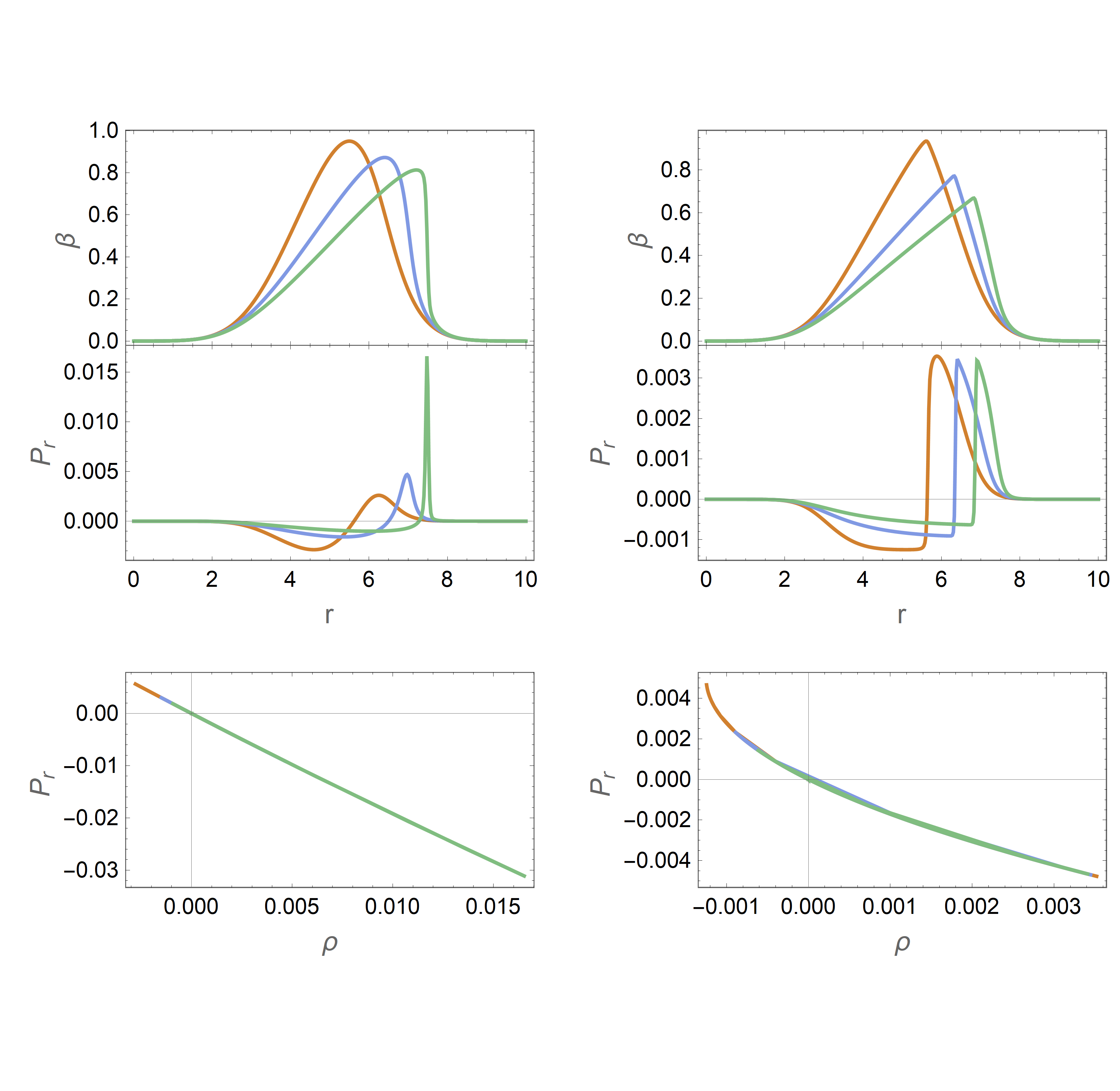}
    \caption{Non-Stationary case 2. Each curve corresponds to times $t=1$ (orange), $t=3$ (blue) and $t=5$ (green). The polytrope parameters are $K=1$ (left column), $K=50$ (right column). The solutions are numerically stable around $\epsilon=-0.5$, $\gamma=2$.
}
    \label{fig:nonsta-case3-1}
\end{figure*}
\end{center}

\begin{center}
    \begin{figure*}
    \centering
    \includegraphics[scale=0.65]{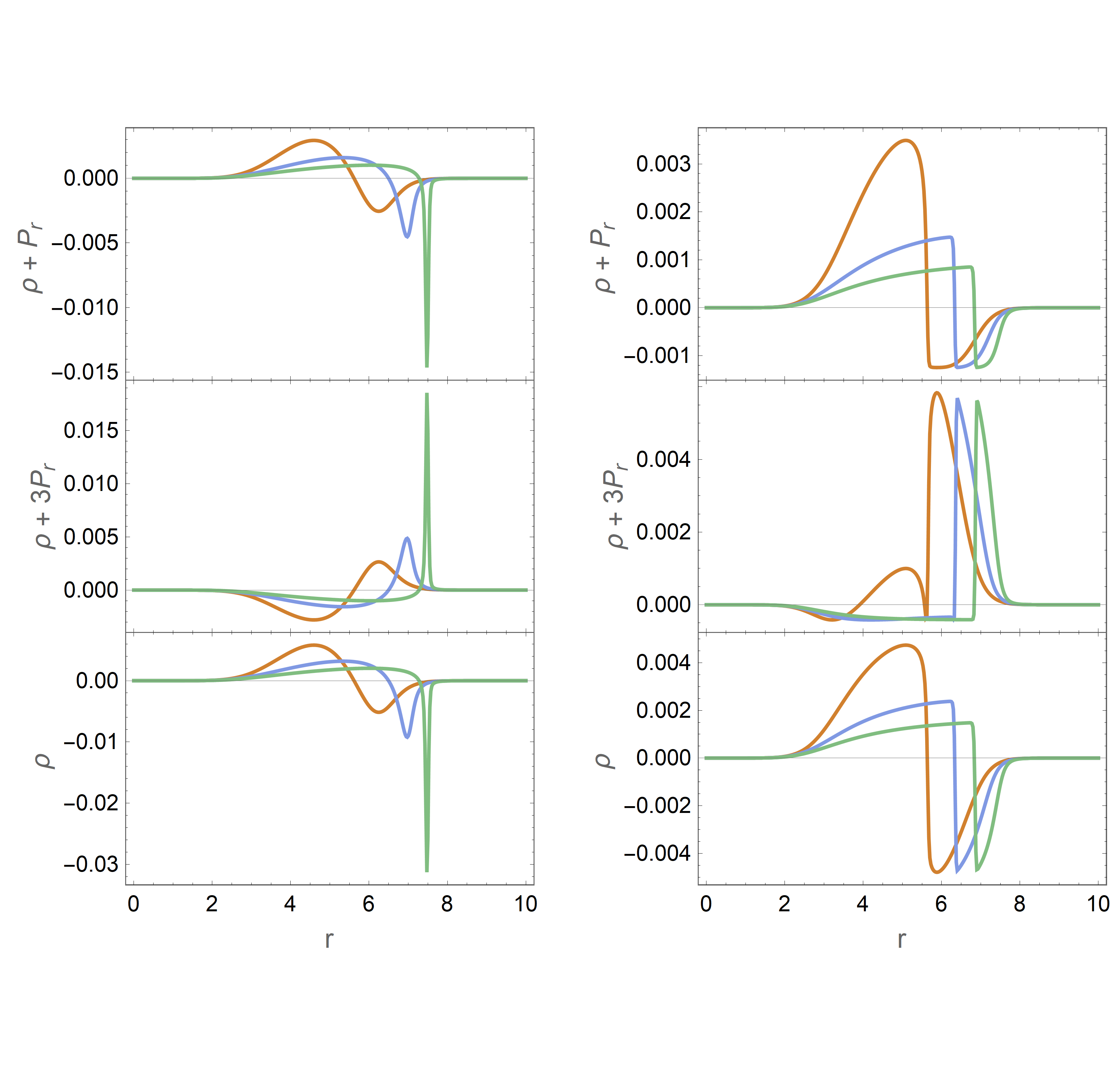}
    \caption{Non-Stationary energy conditions case 2. From the top down we show the NEC, SEC and WEC for $t=1$ (orange), $t=3$ (blue) and $t=5$ (green). The polytrope parameters are $K=1$ (left column), $K=50$ (right column). The solutions are numerically stable around $\epsilon=-0.5$,  $\gamma=2$. 
}
    \label{fig:nonsta-EC-case3-1}
\end{figure*}
\end{center}

Solving for this system we show the results in figs. \ref{fig:nonsta-case3-1} and \ref{fig:nonsta-EC-case3-1}. For the chosen set of parameters, we can observe in fig. \ref{fig:nonsta-case3-1} that the form function $\beta$ is positive and evolves towards a shock wave type behaviour as time progresses. On the other hand, the pressure has regions where it is positive and negative and becomes narrower as time progresses. The empirical equation of state shows a practically linear behaviour for $K=1$ (left column); however, for $K=50$ (right column) the system moves further away from linearity, especially when the density acquires negative values.

In the fig. \ref{fig:nonsta-EC-case3-1} we have the energy conditions for the anisotropic case. We can observe again that for all of them there are regions where they are fulfilled and others where they are not. Moreover, for all cases we notice that energy conditions are bounded and therefore do not have a singular or divergent behaviour. As in the isotropic case, it is possible to use the cosmological constant term to enforce either the weak condition or the strong condition, but not both as mentioned above. We also find that the null condition is violated in some regions.

\subsection{Stationary solutions ($v_s f(r_s)$)}
\begin{center}
    \begin{figure*}
    \centering
    \includegraphics[scale=0.65]{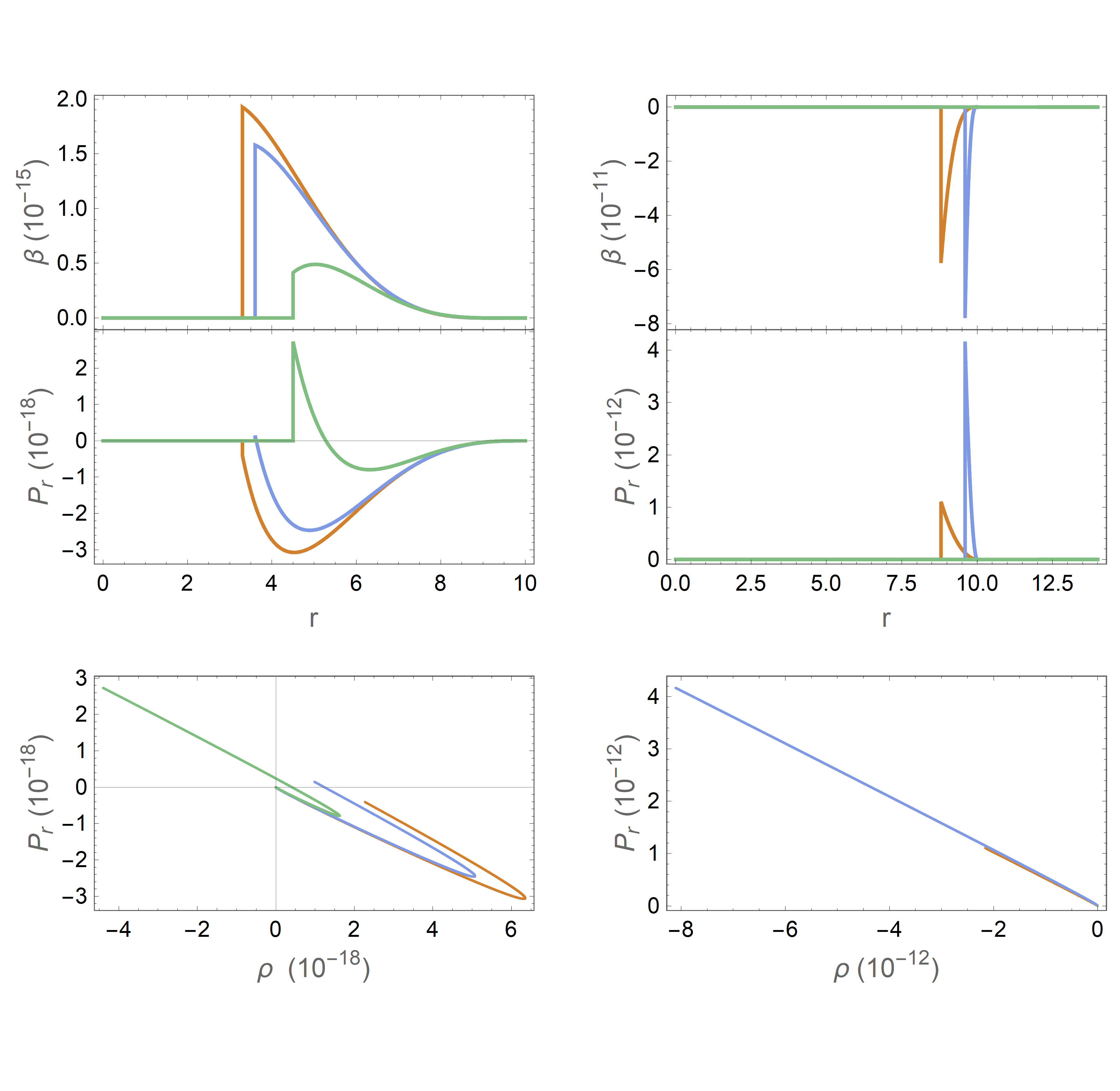}
    \caption{Stationary case 2 for $v_s=0.3$ (left column) and $v_s=0.8$ (right column). The polytrope parameters are $K=100$, $\gamma=2$ and $\epsilon=-0.5$. Each curve corresponds to times $t=11$ (orange), $t=12$ (blue) and $t=15$ (green).
    }
    \label{fig:sta-case2-1a}
\end{figure*}
\end{center}

\begin{center}
    \begin{figure*}
    \centering
    \includegraphics[scale=0.65]{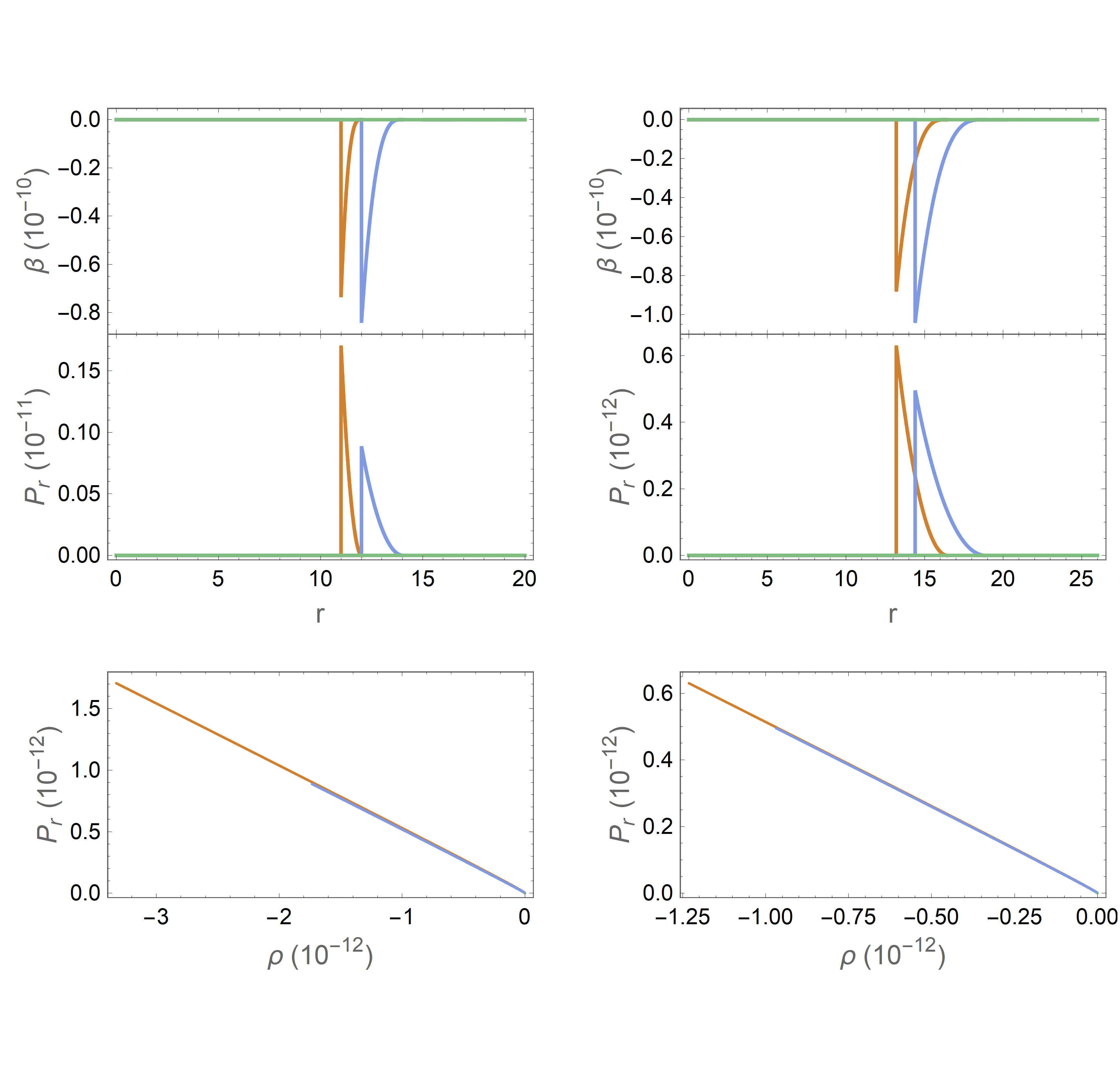}
    \caption{Stationary case 2 for $v_s=1.0$ (left column) and $v_s=1.2$ (right column). The polytrope parameters are $K=100$, $\gamma=2$ and $\epsilon=-0.5$. Each curve corresponds to times $t=11$ (orange), $t=12$ (blue) and $t=15$ (green).
}
    \label{fig:sta-case2-1b}
\end{figure*}
\end{center}

\begin{center}
    \begin{figure*}
    \centering
    \includegraphics[scale=0.65]{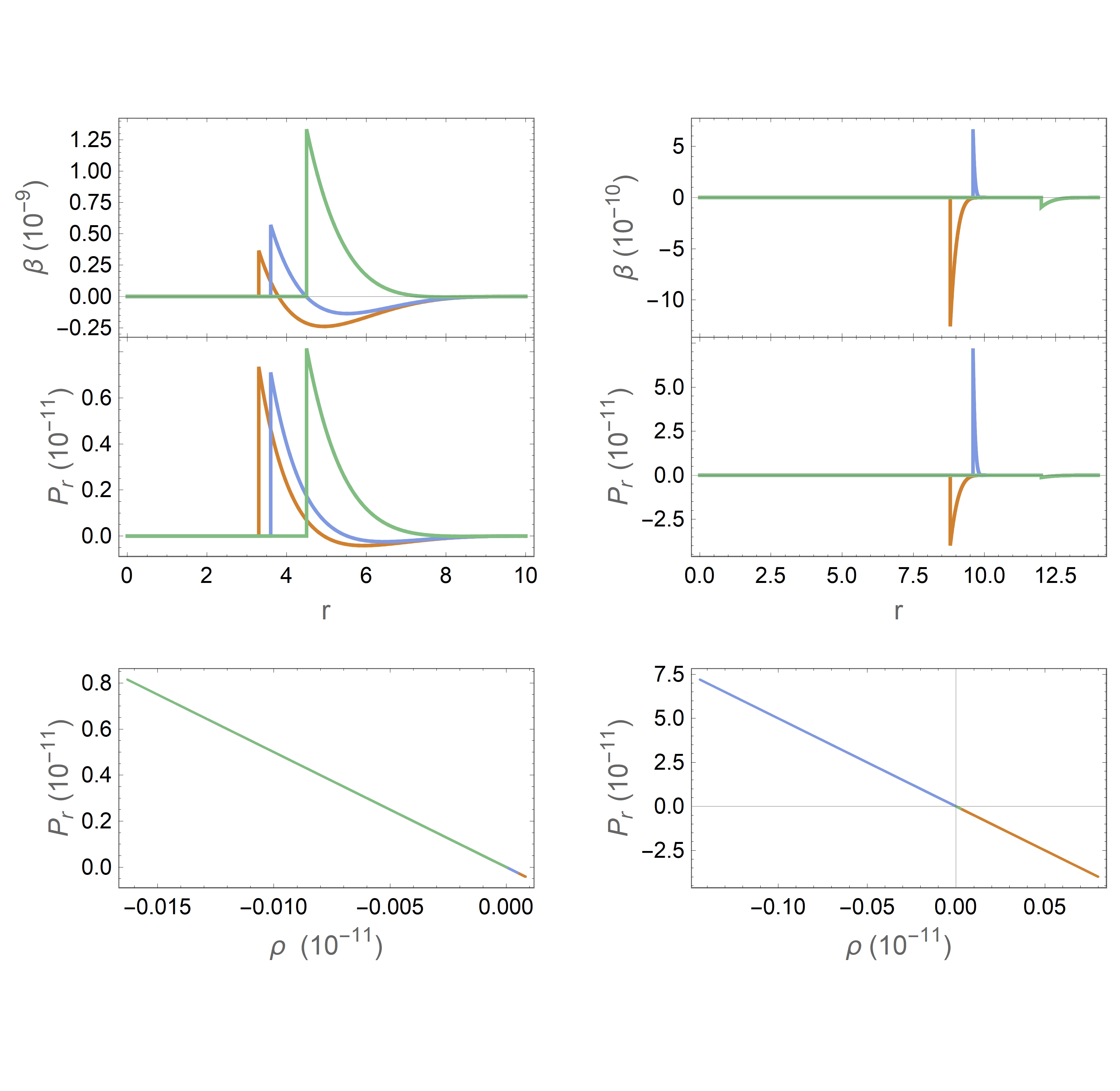}
    \caption{Stationary case 2 for $v_s=0.3$ (left column) and $v_s=0.8$ (right column). The polytrope parameters are $K=100$, $\gamma=2$ and $\epsilon=-50$. Each curve corresponds to times $t=11$ (orange), $t=12$ (blue) and $t=15$ (green).
}
    \label{fig:sta-case2-2a}
\end{figure*}
\end{center}

\begin{center}
    \begin{figure*}
    \centering
    \includegraphics[scale=0.65]{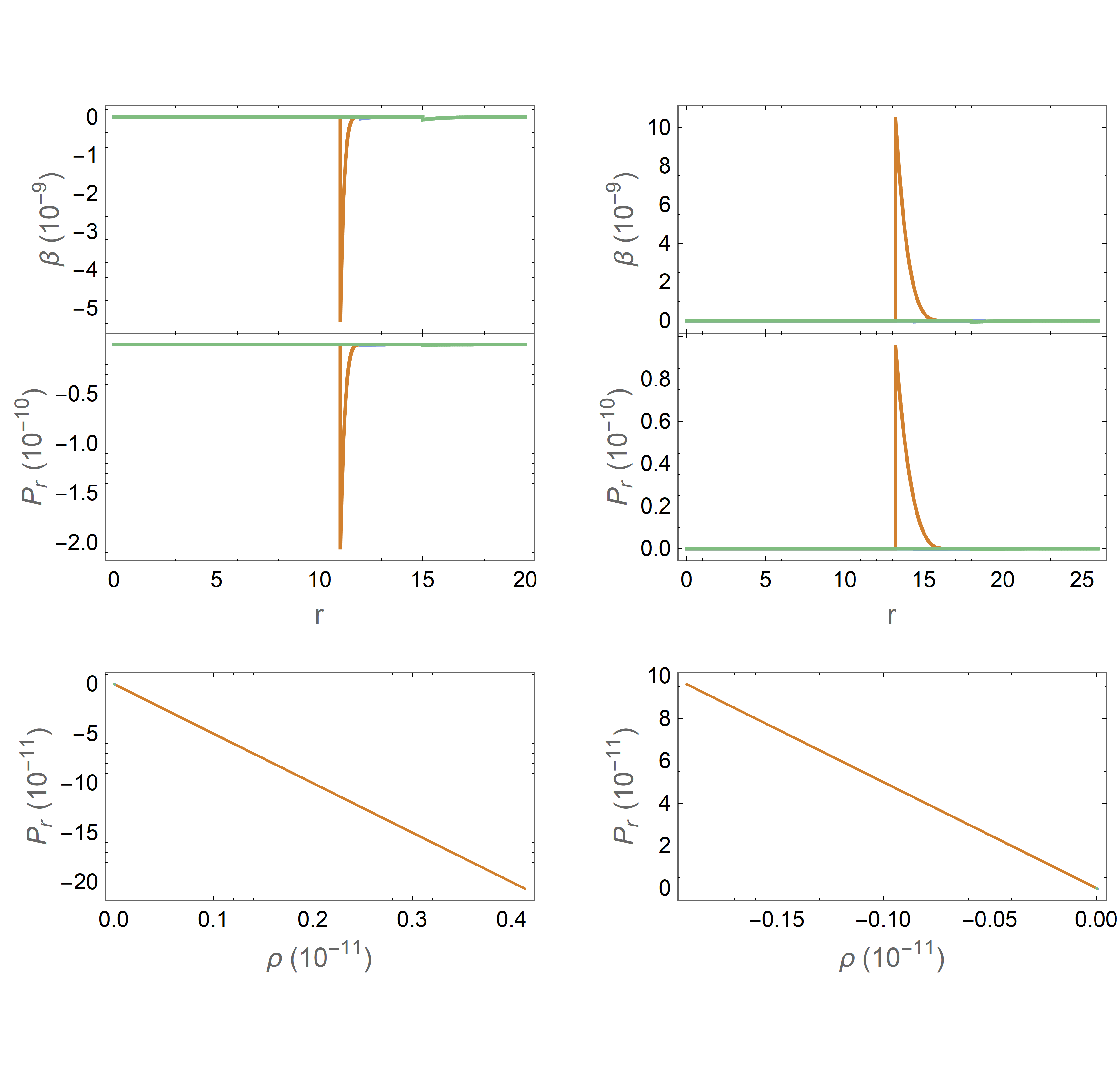}
    \caption{Stationary case 2 for $v_s=1.0$ (left column) and $v_s=1.2$ (right column). The polytrope parameters are $K=100$, $\gamma=2$ and $\epsilon=-50$. Each curve corresponds to times $t=11$ (orange), $t=12$ (blue) and $t=15$ (green).
}
    \label{fig:sta-case2-2b}
\end{figure*}
\end{center}

\begin{center}
    \begin{figure*}
    \centering
    \includegraphics[scale=0.65]{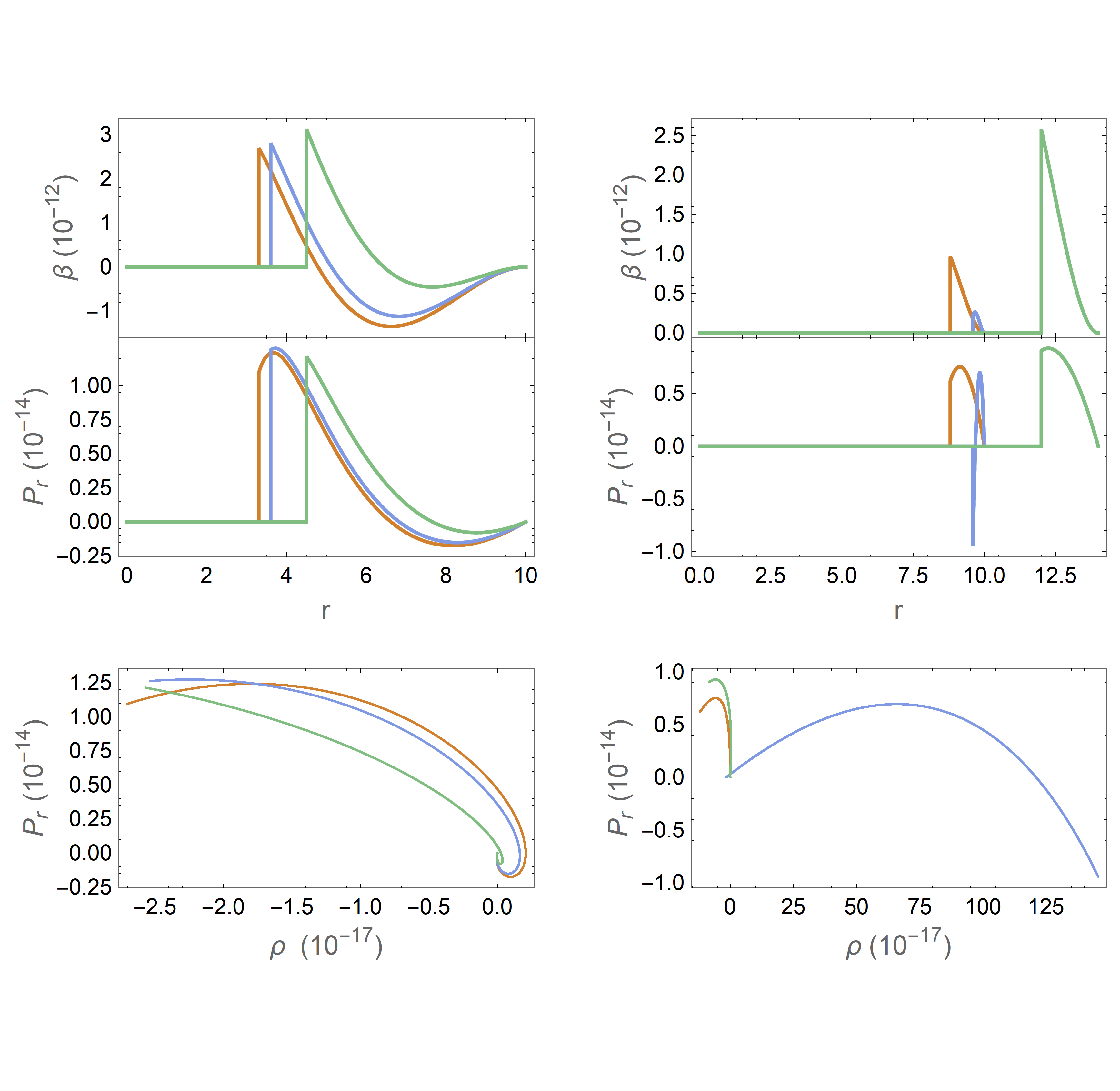}
    \caption{Stationary case 2 for $v_s=0.3$ (left column) and $v_s=0.8$ (right column). The polytrope parameters are $K=100$, $\gamma=2$ and $\epsilon=-10^{-17}$. Each curve corresponds to times $t=11$ (orange), $t=12$ (blue) and $t=15$ (green).
}
    \label{fig:sta-case2-3a}
\end{figure*}
\end{center}

\begin{center}
    \begin{figure*}
    \centering
    \includegraphics[scale=0.65]{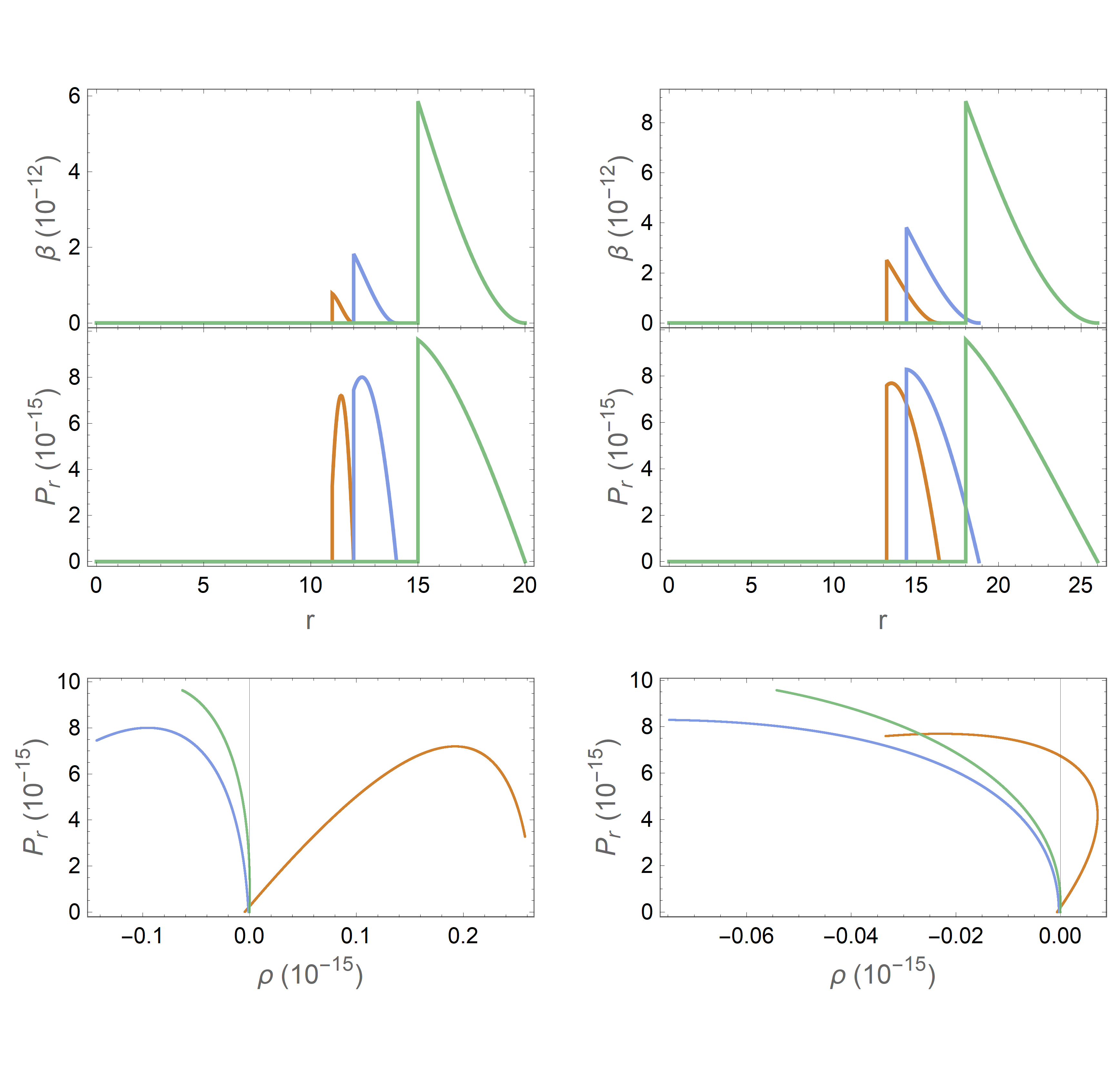}
    \caption{Stationary case 2 for $v_s=1.0$ (left column) and $v_s=1.2$ (right column). The polytrope parameters are $K=100$, $\gamma=2$ and $\epsilon=-10^{-17}$. Each curve corresponds to times $t=11$ (orange), $t=12$ (blue) and $t=15$ (green).
}
    \label{fig:sta-case2-3b}
\end{figure*}
\end{center}

Now we study the traveling wave behaviour of the anisotropic system. In figs. \ref{fig:sta-case2-1a} -- \ref{fig:sta-case2-3b} we show the stationary solutions of $\beta$, pressure $p_r$ and the empirical equation of state $p_r(\rho)$. These are shown for different velocities, from $v_s=0.3$ (subluminal) to $v_s=1.2$ (superluminal). Different sets of parameters have been chosen for the equation of state so that the behaviour of the system in different regimes of the fluid under consideration can be observed.

In general, the form function $\beta$ displaces its origin with the velocity. For $\epsilon=-0.5$ in (\ref{eos}), it can be seen in figs. \ref{fig:sta-case2-1a} and \ref{fig:sta-case2-1b} that $\beta$ is decreasing as time elapses for $v_s=0.3$. However, for $v_s=0.8,1.0,1.2$, we observe that beta is negative and its value for $t=15$ is null while for intermediate times an increase in the amplitude of the form function is obtained. It is worth mentioning that as vs get larger, beta gets narrower, which affects the shape of the bubble as can be deduced from (\ref{expansion}). In the case of radial pressure $p_r$, we find that there are regions where it is positive and others where it is negative. We can also see that as $v_s$ increases, the pressure becomes narrower in accordance with $\beta$. Next, we see the equation of state which produces essentially a linear relationship between $p_r$ and $\rho$ except for $v_s=0.3$ where some non-linearity can be observed.

The analysis for figs. \ref{fig:sta-case2-2a}--\ref{fig:sta-case2-3b} is similar to the previous paragraph. However, it is worth noting that for $\epsilon=-10^{-17}$ you have a completely non--linear equation of state because both terms in (\ref{eos}) have comparable order of magnitude.

\begin{center}
    \begin{figure*}
    \centering
    \includegraphics[scale=0.7]{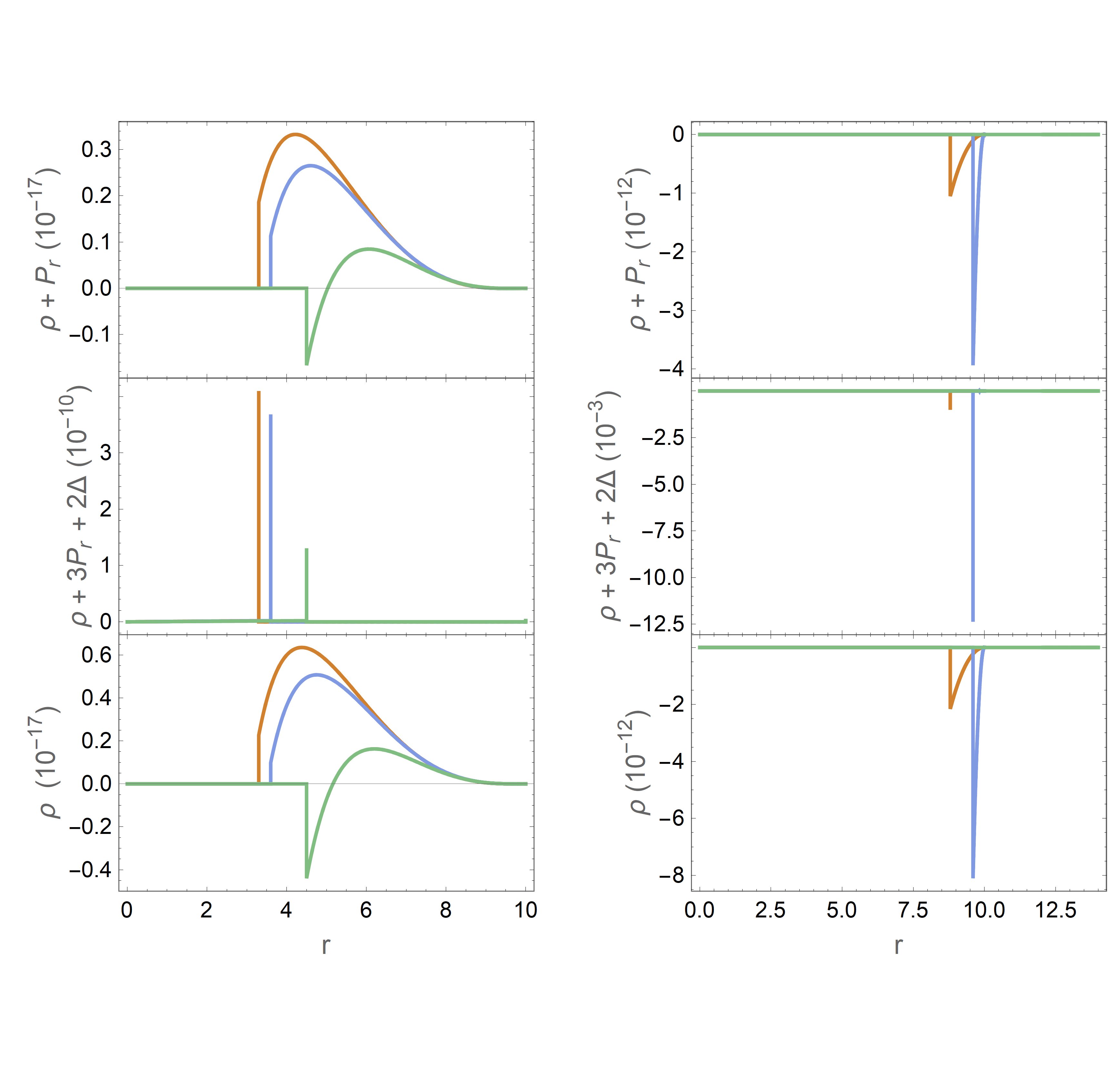}
    \caption{Stationary case 2. From the top down we show the NEC, SEC and WEC for $t=11$ (orange), $t=12$ (blue) and $t=15$ (green). Left column $v_s=0.3$ and right column $v_s=0.8$. The polytrope parameters are $K=100$, $\gamma=2$ and $\epsilon=-0.5$.
}
    \label{fig:case2-EC-1a}
\end{figure*}
\end{center}

\begin{center}
    \begin{figure*}
    \centering
    \includegraphics[scale=0.7]{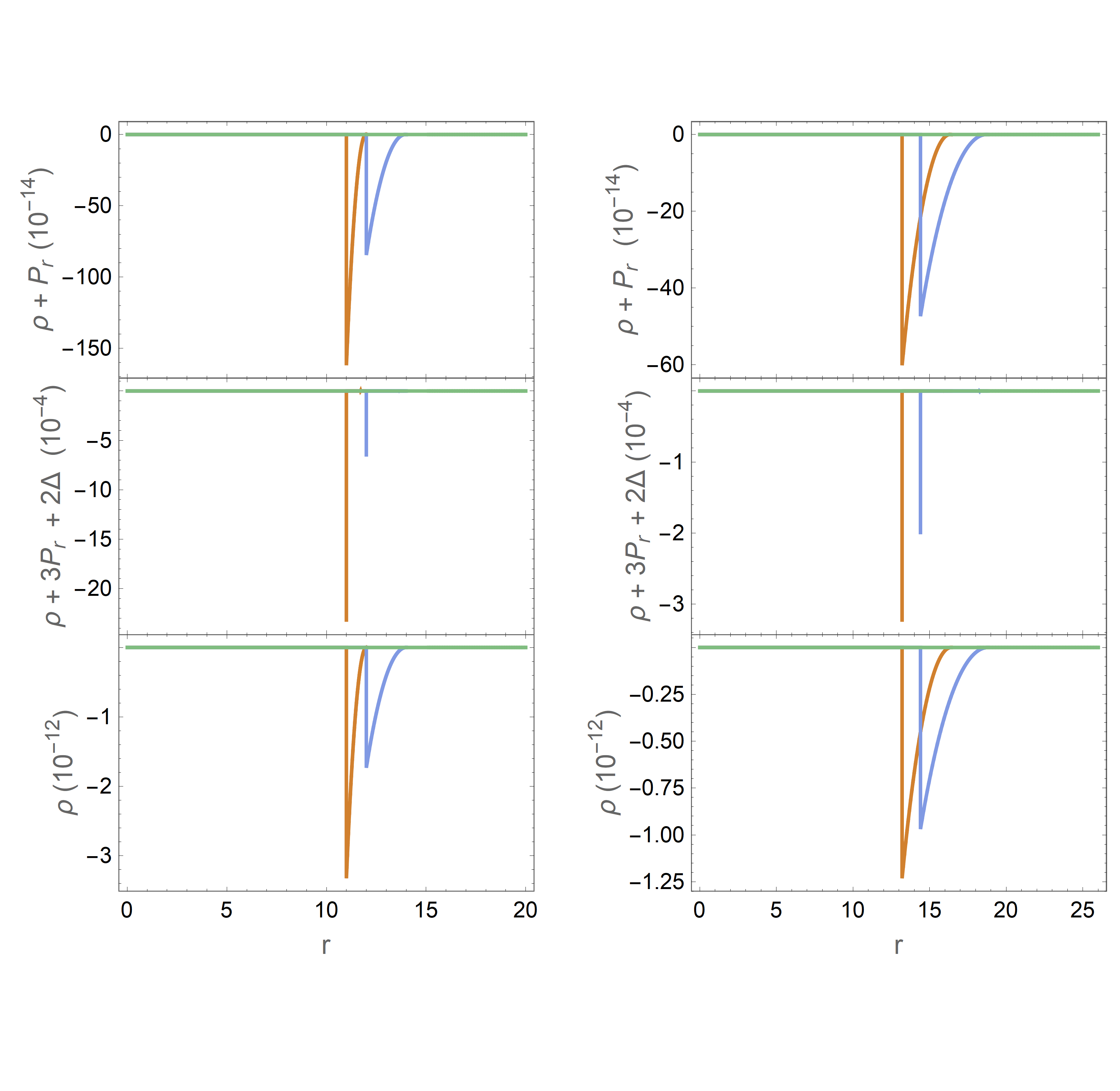}
    \caption{Stationary case 2. From the top down we show the NEC, SEC and WEC for $t=11$ (orange), $t=12$ (blue) and $t=15$ (green). Left column $v_s=1.0$ and right column $v_s=1.2$. The polytrope parameters are $K=100$, $\gamma=2$ and $\epsilon=-0.5$.
}
    \label{fig:case2-EC-1b}
\end{figure*}
\end{center}

As in case 1, it is observed in figs. \ref{fig:case2-EC-1a} and \ref{fig:case2-EC-1b} that the energy conditions have regions where it is positive and others where it is negative. Again, it is possible to use the cosmological constant term to fix either WEC or SEC and the interplay between them is observed where if one is fulfilled, the other is necessarily worsened.

One aspect worth noting is what happens with the weak energy condition which corresponds to the energy density. We can observe that the energy density takes positive or negative values in various regions of the domain. However, we notice that in some cases there are time values where the density is completely positive. This behaviour is then partially lost, which may suggest some kind of instability. 

\begin{center}
    \begin{figure*}
    \centering
    \includegraphics[scale=0.7]{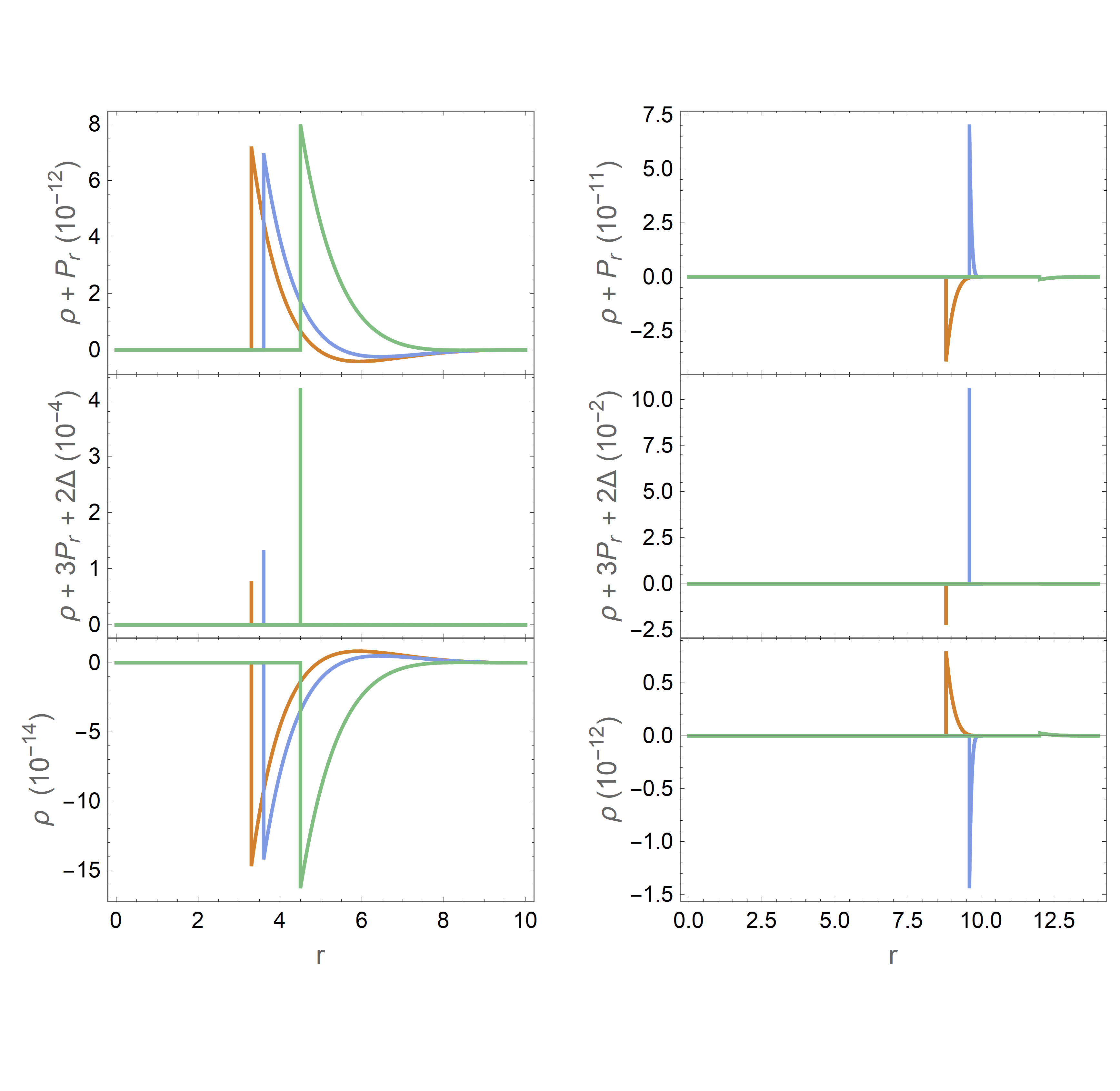}
    \caption{Stationary case 2. From the top down we show the NEC, SEC and WEC for $t=11$ (orange), $t=12$ (blue) and $t=15$ (green). Left column $v_s=0.3$ and right column $v_s=0.8$. The polytrope parameters are $K=100$, $\gamma=2$ and $\epsilon=-50$.
}
    \label{fig:case2-EC-2a}
\end{figure*}
\end{center}

\begin{center}
    \begin{figure*}
    \centering
    \includegraphics[scale=0.7]{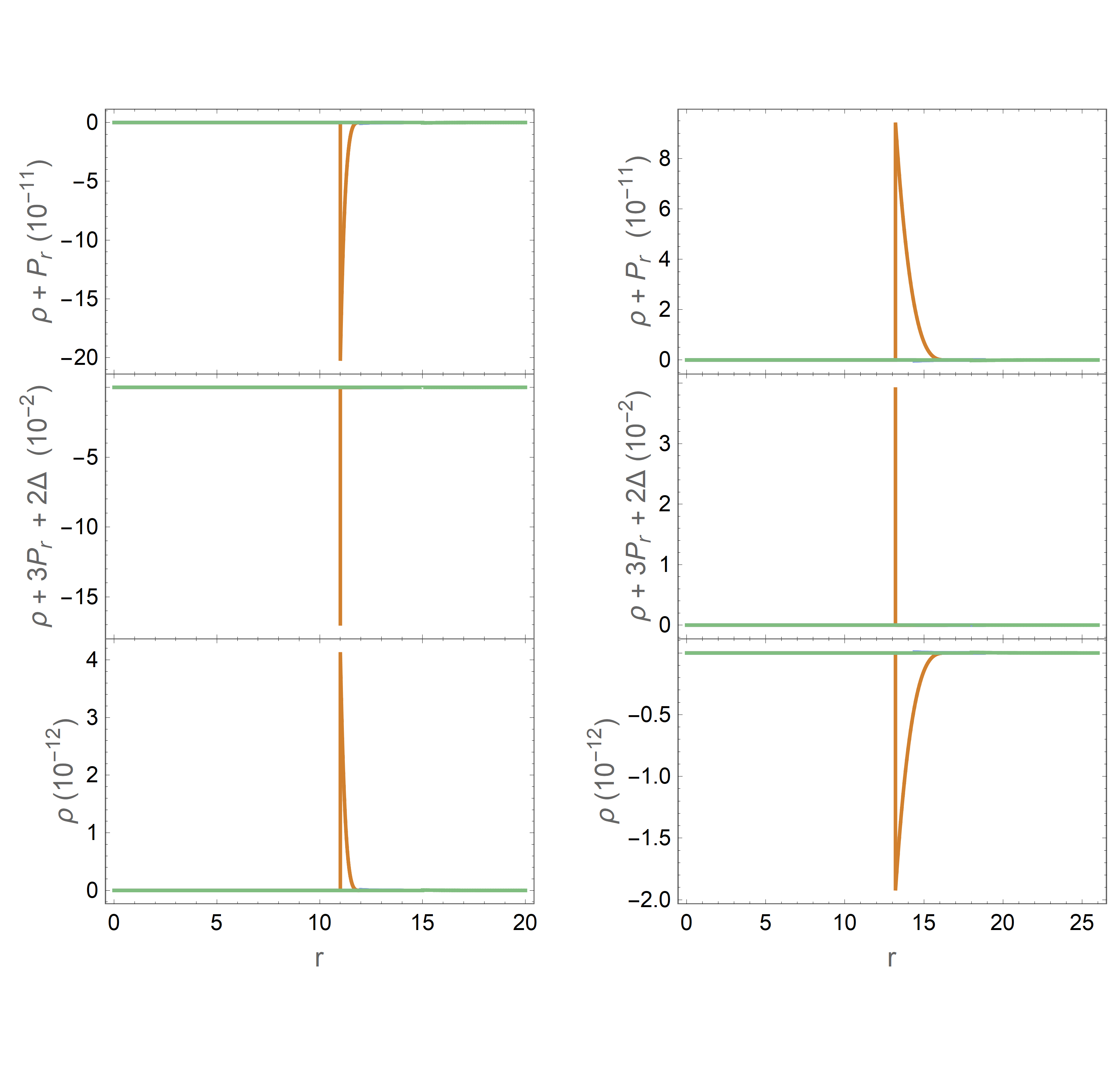}
    \caption{Stationary case 2. From the top down we show the NEC, SEC and WEC for $t=11$ (orange), $t=12$ (blue) and $t=15$ (green). Left column $v_s=1.0$ and right column $v_s=1.2$. The polytrope parameters are $K=100$, $\gamma=2$ and $\epsilon=-50$.
}
    \label{fig:case2-EC-2b}
\end{figure*}
\end{center}

\begin{center}
    \begin{figure*}
    \centering
    \includegraphics[scale=0.7]{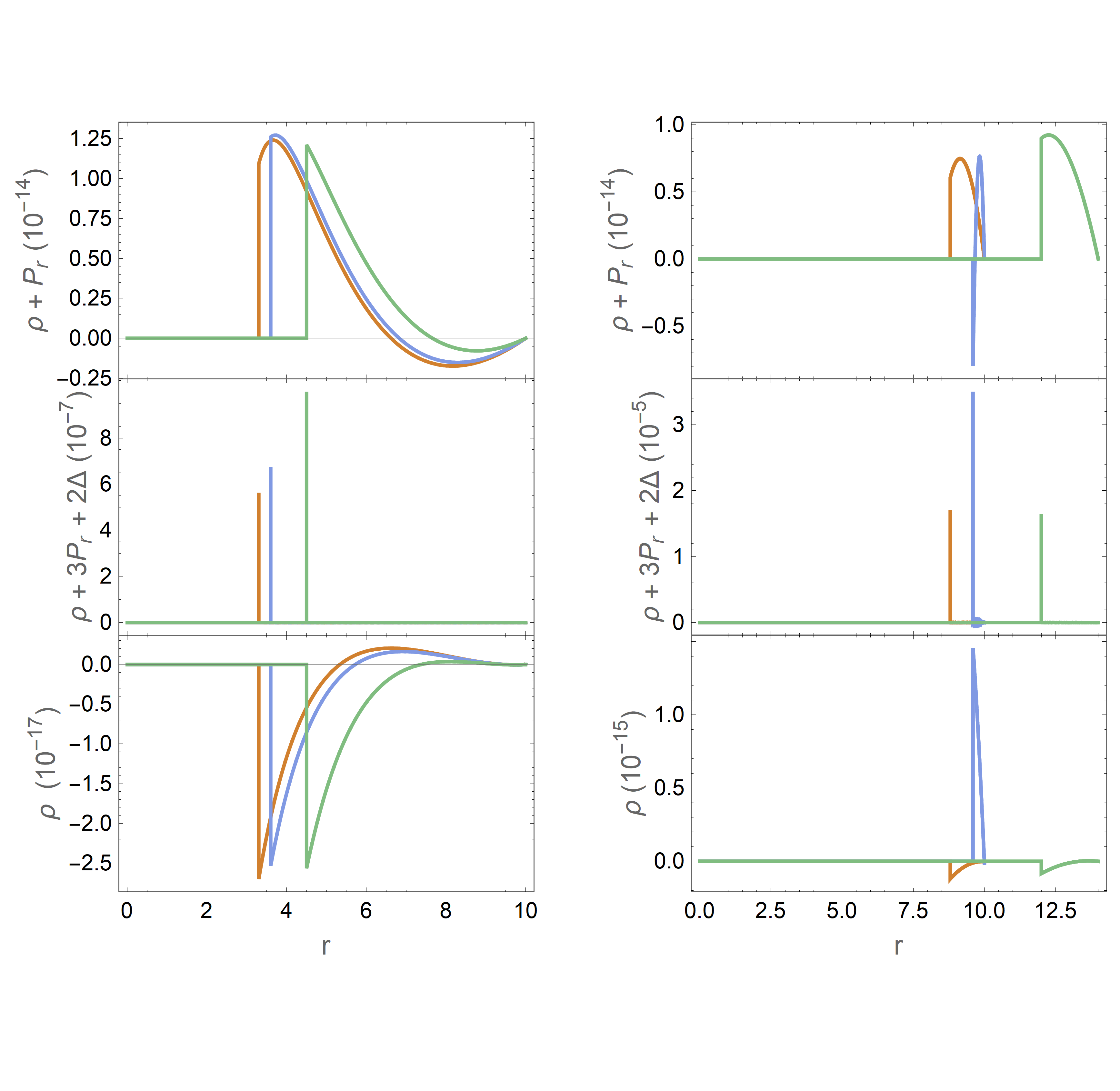}
    \caption{Stationary case 2. From the top down we show the NEC, SEC and WEC for $t=11$ (orange), $t=12$ (blue) and $t=15$ (green). Left column $v_s=0.3$ and right column $v_s=0.8$. The polytrope parameters are $K=100$, $\gamma=2$ and $\epsilon=-10^{-17}$.
}
    \label{fig:case2-EC-3a}
\end{figure*}
\end{center}

\begin{center}
    \begin{figure*}
    \centering
    \includegraphics[scale=0.7]{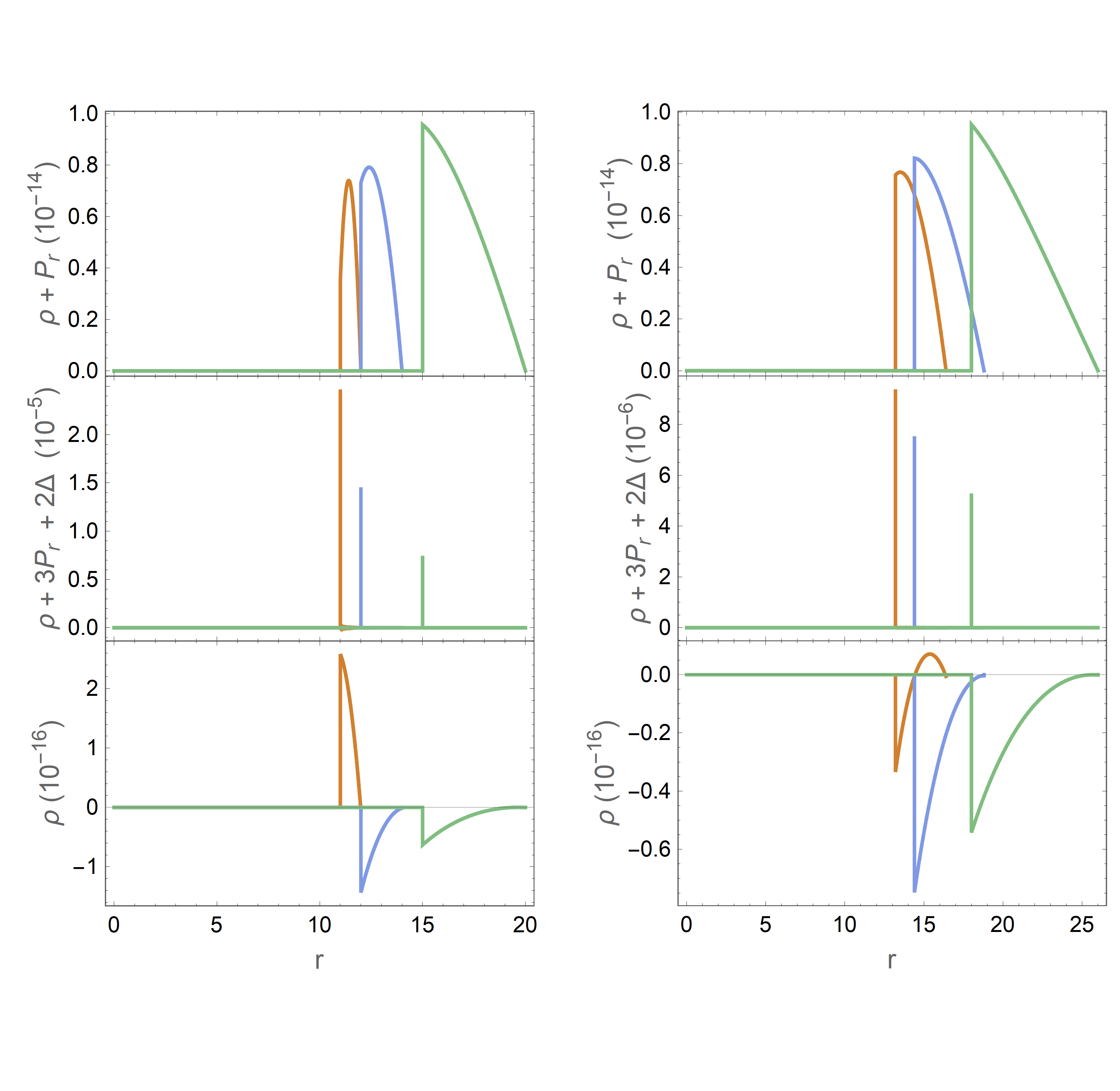}
    \caption{Stationary case 2. From the top down we show the NEC, SEC and WEC for $t=11$ (orange), $t=12$ (blue) and $t=15$ (green). Left column $v_s=1.0$ and right column $v_s=1.2$. The polytrope parameters are $K=100$, $\gamma=2$ and $\epsilon=-10^{-17}$.
}
    \label{fig:case2-EC-3b}
\end{figure*}
\end{center}

\section{Warp density energy requirements}\label{sec6}
Using the results obtained, it is natural to investigate how the mass is described by the configurations of matter used. 
We are interested in determine to what extend the system violates or not the energy conditions, specifically the local weak energy condition. To this end we begin from the relation,
\begin{equation}
    T^{0}_{\;\;0}=  \frac{G_{\;\;0}^0}{8 \pi}  \;,
\end{equation}
which gives an expression for the density 
\begin{equation}\label{rho_warp}
\rho_{\text{warp}} \equiv \rho  = \frac{\beta}{8 \pi r^2} \left(\beta+2r \de{\beta}{r}\right) - \rho_\Lambda \;.
\end{equation}
Next, we can obtain a numerical form of (\ref{rho_warp}) from the numerical solutions obtained for $\beta$. 
We use the ``volume integral
quantifier" proposed by Visser in \cite{sayan:2004}, which amounts to calculating a definite integral for the
relevant coordinate domains.
\begin{eqnarray}
    M_{\text{warp}} = \int_{\mathcal{V}} \rho_{\text{warp}}
    = 4 \pi \int\!\! dr \;  r^2 T^0_{\;\;0} = \frac{1}{2} \int dr \; r^2 G^0_{\;\;0} \;.
\end{eqnarray}
In this way we can quantify the amount of violation or not of the energy density to the extent in which these integrals can become positive or negative for several defined times.

\begin{center}
    \begin{figure*}[ht!]
    \centering
    \includegraphics[scale=2.0]{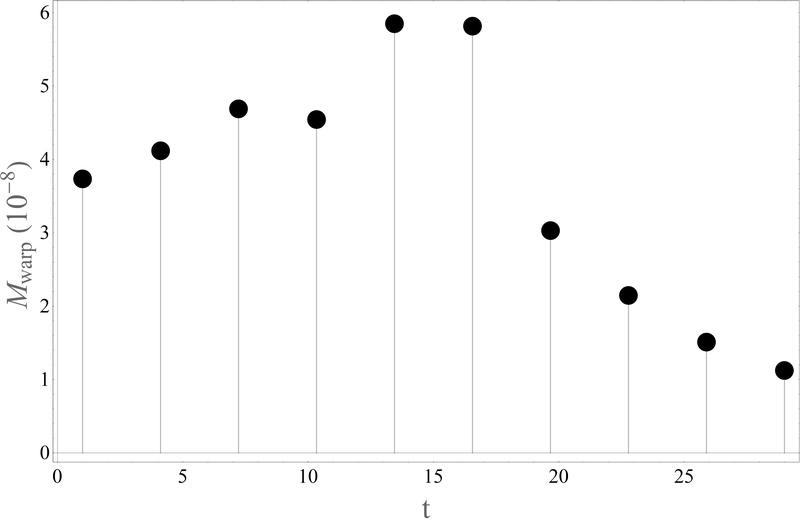}
    \caption{Numerical estimation of $M_{\text{warp}}$ using a ``Volume integral quantifier" integration for several times. The estimation is for stationary case 1 with $v_s=0.3$.}
    \label{fig:warp_mass_case1_sta}
\end{figure*}
\end{center}

\begin{center}
    \begin{figure*}[ht!]
    \centering
    \includegraphics[scale=2.0]{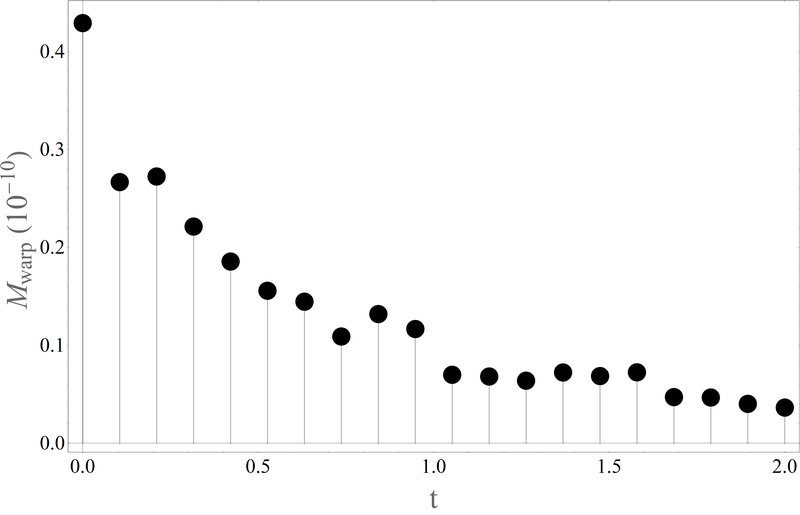}
    \caption{Numerical estimation of $M_{\text{warp}}$ using a ``Volume integral quantifier" integration for several times. The estimation is for non--stationary case 1.}
    \label{fig:warp_mass_case1_nonsta}
\end{figure*}
\end{center}

\begin{center}
    \begin{figure*}[ht!]
    \centering
    \includegraphics[scale=2.0]{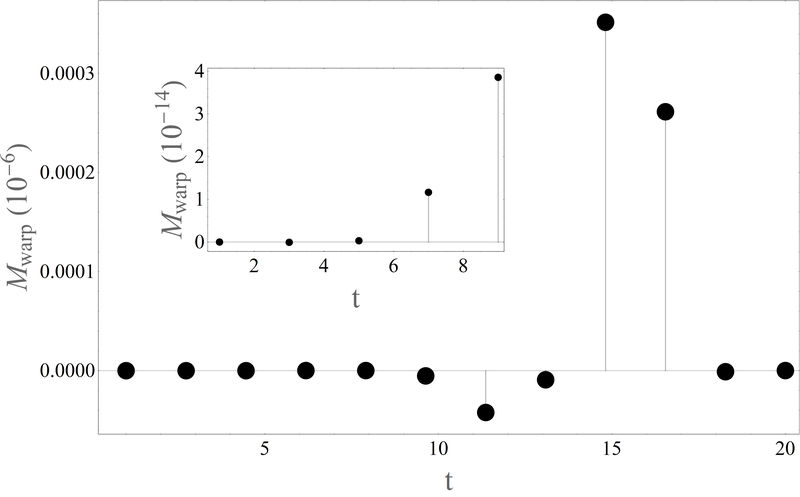}
    \caption{Numerical estimation of $M_{\text{warp}}$ using a ``Volume integral quantifier" integration for several times. The estimation is for stationary case 2 with $v_s=1.0$, $K=0.6$ and $\epsilon=-0.5$.}
    \label{fig:warp_mass_case3_sta}
\end{figure*}
\end{center}

\begin{center}
    \begin{figure*}[ht!]
    \centering
    \includegraphics[scale=2.0]{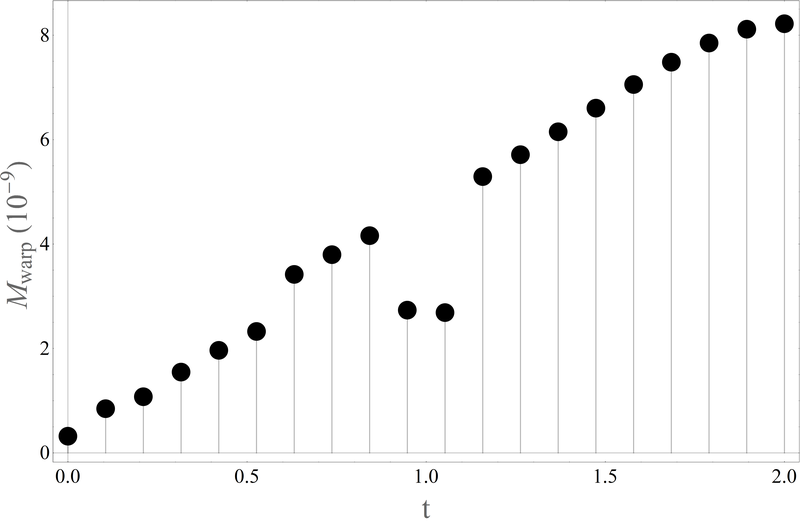}
    \caption{Numerical estimation of $M_{\text{warp}}$ using a ``Volume integral quantifier" integration for several times. The estimation is for non--stationary case 2 with $K=0.6$ and $\epsilon=-0.5$.}
    \label{fig:warp_mass_case3_nonsta}
\end{figure*}
\end{center}

Integrating in the appropriate numerical domain $\mathcal{D}$ given in our solutions $\beta$ with $\rho_\Lambda=0$, we obtain the behaviour for the analysed cases using several values of $t$ which can be seen in Figs. \ref{fig:warp_mass_case1_sta}--\ref{fig:warp_mass_case3_nonsta}. We note first of all that for both examples taken from case 1 we obtain positive mass values throughout the whole time progression. The same is true for the non-stationary case 2. On the other hand, for the stationary case 2 we find some negative mass values. However, most of the points are positive. This result is remarkable and contrasts with the widely debated original Alcubierre's warp.
%We notice that the amount of mass becomes steadily smaller and smaller in an oscillatory manner.
Apart from observing a large sample of positive values for the required masses of the warp, we also notice that they are of a rather low order of magnitude compared to what has been reported in previous work.

For instance if we pick the mass point of case 1, fig. \ref{fig:warp_mass_case1_nonsta} associated with time $t=2.0$ which has order of magnitude $10^{-11}$ it corresponds to masses of order $10^{-16}$ kg. Compared to some familiar Newtonian masses, it is equivalent to $10^{-7}$ the mass of the moon, or $57$ times the mass of mount Everest.

It is expected that since the null energy conditions demands in equation (\ref{null_condition_beta}) the $\beta$ function to diminish in time, the values of $M_{\text{warp}}$ will tend to zero. An interesting aspect of this solutions is the oscillating behavior shown, so the quantity $M_{\text{warp}}$ seems oscillates and damps as times evolves. A remarkable difference occurs in fig. \ref{fig:warp_mass_case3_sta}, where although the warp mass stays positive it seems to increase in time. We believe that because the method for finding these solutions is unstable for long times, we were not able to find the expected decaying regions.

%-----------------------------------------------------------------------------

\section{Final Remarks}
In this paper we have explored various solutions to a spherically symmetric metric describing a warp drive \cite{abellan:2023}. For them we have considered as a source of matter a generic fluid which admits anisotropy and cosmological constant. Once the Einstein equations for the system under study have been established, we realise that the anisotropic fluid is the most general that we can describe with the proposed metric. The metric used here has obvious advantages over the traditional metric originally proposed by Alcubierre. This is due to the incorporation of spherical symmetry in describing the residual flat space that remains in any warp metric. Doing so allows a much more neat and ready to be solved set of equations.

We have rewritten the set of equations found to accommodate travelling wave solutions. We found this useful for several reasons. Firstly, it is this type of solution that was originally proposed by Alcubierre when he gave an explicit form of the metric, so we wanted to get as close as possible to this approach in order to be able to compare. Secondly, the set of differential equations becomes dependent on a single variable and this greatly simplifies their solution. And thirdly, as a consequence of the previous point, we avoid the use of initial conditions, which can add a certain degree of arbitrariness to the solutions obtained. The solutions described in this way can be considered as the stationary regime to the problem posed.

As we have already described, the problem of energy conditions has dominated the discussion regarding the feasibility of warp solutions. For this reason, we include a detailed analysis of the energy conditions from the point of view of both the energy-momentum tensor and the metric coefficients. Trivially the dominant energy condition is always satisfied. On the other hand the null energy condition imposes that the $\beta$ form function must decrease in time or at least remain unchanged. An interesting aspect occurs when including the cosmological constant term as a material source. We find that there is a trade-off between weak and strong energy conditions. We could use the cosmological constant to fix one of them but then we would necessarily make the other worse. This relationship is very interesting and worth further exploration. Another aspect worked on is the incorporation of the zero expansion condition in the system. This produces a set of equations with little room for experimentation. However, we believe it is worth taking into account for future work.

The equations found were solved using numerical methods. For this purpose, we considered both stationary solutions using the above-mentioned description of travelling patterns, and complete solutions without imposing any restrictions. Both types of solutions were applied to isotropic and anisotropic fluids. In general, it is observed that for certain time instants there are spatial regions where the energy conditions are satisfied. In the isotropic case it is interesting to observe the type of empirical equation that appears, reflecting a multivalued behaviour which may suggest some type of fluid that allows the occurrence of phase transitions. The characterisation of these fluids is a very interesting aspect but clearly beyond the scope of the present work. Regarding the energy conditions, we can observe that all of them show plots with bounded values. This absence of divergence is a good sign and motivates us to continue exploring mechanisms to satisfy the conditions in a general way. In this work we explore the effect of adding a cosmological constant-type term to the momentum energy tensor. This allows us to correct the behaviour of either the weak or the strong energy condition but not both. This trade-off between the two conditions seems to be a general property. We can say that exploring different kinds of fluids opens up new possibilities and may provide a way to deal with the constraints imposed by the energy conditions. 

With respect to the $\beta$ form function, in general it is observed that its amplitude is decreasing. However, there are some of the simulated cases in which the opposite behaviour appears and the amplitude increases. This seems to suggest a certain instability in the solutions. Something similar can be observed for the radial pressure, which in most cases is at least partially positive and only in a few cases completely negative.

Finally, we calculate the value of the total warp mass using the results found in the numerical simulations. This is useful because it acts as an average indicator of the amount of total energy needed to sustain the warp structure. We were able to find values in the parameters that make the integrated energy density positive at all instants of the system's evolution. In most cases we noticed that these integrated values become smaller and smaller with time. This may support the argument of some instability in the solutions, although we believe that more tests are needed to have a conclusive result. However, the fact of finding solutions that produce positive mass values is remarkable and we believe that further study of this system will allow us to understand the fundamental characteristics of warp drives and also to understand why solutions like Alcubierre's behave as they do.

In summary, we believe that research related to the warp spherical metric should be continued as it presents an efficient way to study analytically and numerically the problems associated with the warp drive phenomenon. Furthermore, it is crucial to continue studying the material content that supports the warp geometry, distinct and more complex matter configurations, different metric realizations that could include dissipation, heat flux and electromagnetic fields can be studied, in this way to pointing to tailor feasible space-times with these configurations, at least at a theoretical level.

\bibliography{bibliocases.bib}% Produces the bibliography via BibTeX.

% common bib file
%% if required, the content of .bbl file can be included here once bbl is generated
%%\input sn-article.bbl

%% Default %%
%%\input sn-sample-bib.tex%

\end{document}